\documentclass[conference]{IEEEtran}

\usepackage{balance}

\IEEEoverridecommandlockouts 
\usepackage[dvips]{graphicx}
\usepackage{algorithmic}
\usepackage{algorithm}
\usepackage{listings}
\usepackage[OT4,T1]{fontenc}
\usepackage[cmex10]{amsmath}
\usepackage{url}
\usepackage{multirow}
\usepackage{tabularx}
\usepackage{caption}
\usepackage{subcaption}
\usepackage[hidelinks]{hyperref}
\usepackage{booktabs}
\usepackage{mathrsfs}
\usepackage{microtype}
\usepackage{enumitem}

\usepackage{xcolor}

\title{Dynamic communication topologies for distributed heuristics in energy system optimization algorithms}
\author{
\IEEEauthorblockN{Stefanie Holly}
\IEEEauthorblockA{
R\&D Division Energy\\
OFFIS - Institute for Information Technology\\
Escherweg 2, 26121 Oldenburg, Germany\\
Email: stefanie.holly@offis.de}
\and
\IEEEauthorblockN{Astrid Nieße}
\IEEEauthorblockA{
R\&D Division Energy \\
OFFIS - Institute for Information Technology\\
Escherweg 2, 26121 Oldenburg, Germany\\
Email: astrid.niesse@offis.de}
}

\begin{document}
\maketitle              
\begin{abstract}
The communication topology is an essential aspect in designing distributed optimization heuristics. It can influence the exploration and exploitation of the search space and thus the optimization performance in terms of solution quality, convergence speed and collaboration costs – relevant aspects for applications operating critical infrastructure in energy systems. In this work, we present an approach for adapting the communication topology during runtime, based on the principles of simulated annealing. We compare the approach to common static topologies regarding the performance of an exemplary distributed optimization heuristic. Finally, we investigate the correlations between ﬁtness landscape properties and deﬁned performance metrics.\let\thefootnote\relax\footnote{Preprint. In press}
\end{abstract}
\section{Introduction}
\IEEEoverridecommandlockouts\IEEEPARstart{D}{istributed} heuristics are a promising field for current and future energy systems control and optimization tasks, 
and have been designed and evaluated in recent years on agent-based systems \cite{Vrba2014} \cite{Ramchurn2011} \cite{sonnenschein2015}.
While conventional control systems -- centralized or hierarchical in their control paradigm -- perfectly fit to centralized generation and transmission systems, distributed renewable energy systems show properties that promote the application of distributed optimization systems: First, future energy systems can be regarded as complex systems of systems, sometimes framed as cyber-physical multi-energy systems, coupling communication systems, power, heat and gas systems. The resulting complexity of the solution space is the main motivation for heuristic distributed control and optimization \cite{Niesse2013a}. 
Second, data availability as needed for centralized control typically is not given for end-user scenarios for privacy or regulatory reasons.
\par Distributed control and optimization systems often involve multiple energy units that decide locally and communicate with each other to solve global problems.
For instance, software agents can represent flexible energy loads that cooperatively aggregate flexibility to provide load dispatch options for balancing markets or congestion management \cite{HU2015} \cite{Lai2019} \cite{Niesse2018}.
\par One major design aspect is the communication topology. This topology - usually modeled as a graph - determines which units exchange data directly.
In energy system applications, the communication topology of multi-agent systems is often defined based on the topology of the underlying power grid (see e.g.\ \cite{Kok2005} and \cite{Lehnhoff2011}). 
This approach is limited to static topologies, and not reflecting algorithmic aspects. 
Considerable research has been conducted on distributed optimization in the power grid, in the area of control theory, where systematic mathematical approaches are used to design distributed controllers \cite{molzahn2017survey}. However, the type of problems that can be solved with such approaches is limited \cite{rizk2018decision}. Distributed energy resources are very heterogeneous regarding forecast precision and flexibility potential.
Agents have to consider more local constraints and be able to react flexibly to their environment.
Thus, we consider algorithms that provide a framework for negotiation between different agents, but allow flexible local action.
To the best of our knowledge, there has been little research on how the communication topology affects such distributed heuristics or how it can be optimally designed.
\par In \cite{holly2020effects} we showed that different communication topologies have an 
effect on the performance of the reflected algorithm class: 
Highly meshed topologies converged into good solutions reliably and quickly, but increased communication overhead and premature convergence.  
In contrast, results for sparsely meshed topologies were much less reliable. 
In the application domain of energy systems as critical infrastructures, this behavior is highly unwanted.
We presume that dynamically adjusting the topology during runtime leads to a beneficial transition of exploration and exploitation of the search space for distributed heuristics.
\par In this contribution, we evaluate the effect of dynamic communication topology adaptation on a fully distributed optimization heuristic. 
To ensure scientific comprehensibility and reproducibility, standard optimization problems are taken for an extensive analysis of the approach. 
Furthermore, we analyse correlations between the performance of the distributed optimization algorithm and the fitness landscape characteristics of these benchmark functions with both static and dynamic overlay topologies using decision trees.
\par The rest of this contribution is structured as follows: In section \ref{sec:related}, an overview on the topic of communication topologies for distributed heuristics is presented, motivating the research gap.
The dynamic topology adaptation scheme is presented in section \ref{sec:topo_adaptation}. The metrics used for the fitness landscape analysis are presented in section \ref{sec:landscape}. In section \ref{sec:methods} we set the scene for the experimental setup chosen to analyse the relevant correlations, followed by a discussion in section \ref{sec:results}. We conclude our work with an outlook on future research directions.
\section{Communication topologies for distributed heuristics}
\label{sec:related}
Distributed optimization heuristics are closely related to parallel cooperative metaheuristics  \cite{talbi2009metaheuristics}. In both types of heuristics, multiple distributed (meta)heuristics are interconnected and exchange information. Therefore, we consider the studies on the influence of exchange topologies in the area of parallel cooperative metaheuristics as relevant related work.

More precisely, we restrict our scope to asynchronous cooperative search strategies, i.e.\ several solvers run simultaneously (multi-search / distributed on algorithmic level) and cooperate with each other by asynchronously exchanging information.
This type of parallel heuristics originates from the field of parallel computing.
Therefore, the communication topology was mostly designed with respect to the hardware architecture (considering connections between processing units) leading to hypercube, ring or torus topologies \cite{rucinski2010impact}.
For island model heuristics, i.e., heuristics where multiple instances of mostly population-based metaheuristics run in parallel exchanging individuals between their islands, a fully meshed topology is often chosen \cite{crainic2019parallel}.
In addition, information is usually exchanged indirectly via a shared memory. 
\par The effect of communication topologies on the performance of distributed optimization heuristics has been studied especially for the island-model, where the topology is often referred to as the migration topology.
In most works, different topologies for a given parallel heuristic are studied on multiple benchmark problems and the topologies are ranked according to the achieved performance, which may involve different aspects \cite{rucinski2010impact}, 
\cite{hijaze2009investigation},  \cite{hijaze2011distributed}, \cite{sanu2015empirical},  \cite{wang2019empirical}.
Ruci\'{n}ski et al. investigated the effect of different migration topologies, including ring, cartwheel and hypercube topologies, on the performance of two different parallel global optimization algorithms cooperating via the island model \cite{rucinski2010impact}. 
They evaluated different topologies for both heuristics according to the performance obtained.
Since the results varied widely, Ruci\'{n}ski et al. suggested that such studies be conducted in the future for other heuristics and with more problem instances. 
\par Hijaze and Corne \cite{hijaze2009investigation} analyzed how different topologies affect the performance of an asynchronous distributed evolutionary algorithm (EA). 
They evaluated the effect of the topologies on the performance of the algorithm using 30-dimensional target functions (Sphere, Rosenbrock, Schwefel, Rastrigin, Griewank, Ackley \cite{momin2013literature}\cite{li2013benchmark}). The success rate in finding the optimum was similar for all topologies, but better than for the standard single population EA (with equal total population size). 
\par In their follow-up work in \cite{hijaze2011distributed}, they introduced an online adaptation of the migration scheme in which the migration probability was adjusted based on the progress of subpopulations on islands.
With the adaptive scheme, optimal solutions were regularly found in less time and with a higher success rate, suggesting that a balance between exploration and exploitation can be achieved by dynamically adjusting the migration mechanisms. 
\par In \cite{sanu2015empirical}, Sanu and Jeyakumar conducted an empirical analysis on the performance of distributed differential evolution (DE) for varying migration topologies.
They used various topologies (basic ring and ring variants, star, cartwheel, torus and mesh) and multiple benchmark functions (e.g.\ Sphere, Schwefel (1,2,3), Rosenbrock, Rastrigin) to investigate the impact of the topologies on the performance of an island model DE.
They considered not only the convergence speed and solution quality based metrics, but also the computational effort, i.e., the number of function evaluations. 
They concluded that no single topology is suitable for all optimization problems and took a first step towards linking characteristics of the search spaces to the performance of the topologies by roughly categorizing the functions (modality and separability) and assigning the best performing topologies in each case. 
\par The presented research can be summarized as follows: 
First, different communication topologies affect the performance of the various parallel metaheuristics. 
Second, the notion of performance is mainly limited to the achieved solution quality and convergence speed.
Since the studies do not address spatially distributed systems, they usually do not examine the costs of collaboration, especially the resulting message traffic.
Third, the design of communication topologies leads to different balances between exploration and exploitation of the search space. In some cases it is investigated how this balance can be improved by adjusting parameters like migration frequency or migration rate. However, the adaptation of the topology
has not been treated as a distinct research topic for optimization problems with characteristics as be found in energy system applications.
To our knowledge, a systematic approach regarding the above mentioned aspects including an in-depth fitness landscape analysis has not yet been conducted in this field. 

\section{Dynamic Topology Adaptation}
\label{sec:topo_adaptation}
As described in the previous section, the communication topologies of distributed optimization heuristics affect the degree of exploration and exploitation of the search space. Strongly meshed topologies lead to a high amount of information exchange between units (diversification). This leads to a fast convergence but bears the risk of a premature convergence into local optima. In contrast, with sparsely meshed topologies, the individual heuristics can evolve more independently. This leads to an intensified search (exploitation) in some areas of the search space. But in the worst case, these areas can be far away from the global optimum.
\par Many metaheuristics adjust parameters at runtime to allow a transition from exploration to exploitation. An example of this is the adjustment of the temperature parameter $T$ for simulated annealing (SA) \cite{blum2003metaheuristics}. 
\par
\begin{figure}[tbp]
  \centering
  \includegraphics[width=\linewidth]{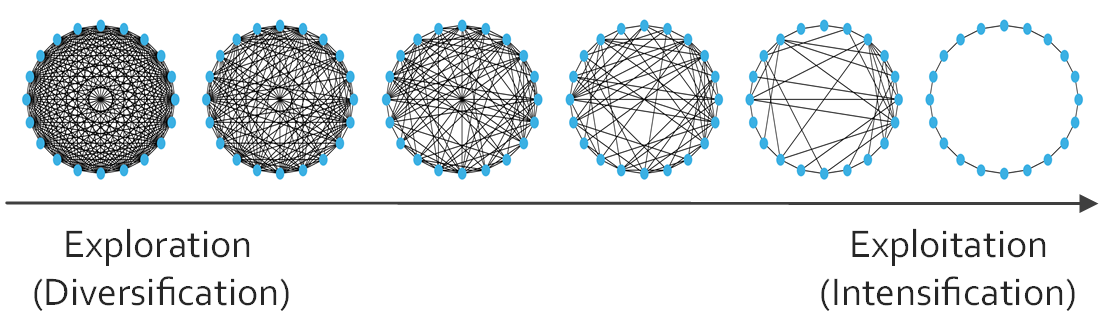}
  \caption{Dynamic topology adaptation}
  \label{fig:explo_to_exploi}
   \vspace{-1.5em}
\end{figure}

 Since we want to achieve the same effect with dynamic topology adjustment, our approach is based on the principles the cooling process in SA. Just as SA starts with a high temperature, we start with a high number of connections in the communication topology. With cooling down, we reduce the number of connections. 
To model the cooling process, SA uses a so-called cooling schedule. We therefore determine a so called "removal schedule". \autoref{tab:removal_sched} shows the details of the specified analogy.
\begin{table*}
\centering
  \caption{Comparison of the modeling of SA cooling and the connection reduction of the communication topology}
  \label{tab:removal_sched}
 \begin{tabular}{p{0.2\linewidth} p{0.34\linewidth}p{0.38\linewidth}}
    \toprule
    & cooling schedule & removal schedule\\
    \midrule
    definition of & temperature for each step of the SA algorithm & number of
edges in the communication topology for each step\\
    initialization parameter & $T_0$: initial temperature& $|E_0|= \delta_0$: initial number of edges\\
    equilibrium state & number of iterations at a temperature & number of local optimizations at a topology configuration\\
    adaptation & cooling: decrease of the
temperature & decrease of edges in the
communication topology\\
  \bottomrule
\end{tabular}
\end{table*}
The approach presented here starts with a fully meshed topology, transitions to small world intermediate stages by removing edges, and ends with a ring to exploit the most promising regions in the solution space. \autoref{fig:explo_to_exploi} illustrates this process.
\par In order to model this transition, more specifications are necessary.
Let $G = (V, E)$ denote the bidirectional graph that represents the communication topology. $V$ is the set of nodes, where each node is assigned to an agent and thus to one part of the distributed solver. $E$ is the set of edges. An edge between two nodes indicates direct communication between the two agents assigned to the nodes.
Agents can pass on information from their neighbors to other neighbors, which means that there is also indirect communication between unconnected nodes. The communication topology thus regulates the information dissemination in the distributed system.
\par A cooling schedule for SA is determined by the initial temperature $T_0$, the equilibrium state, i.e. the criterion that controls when the transition to the next temperature level occurs, and the cooling itself. 
Since the communication topology will start with a fully meshed bidirectional graph, the number of edges is defined as
$|E_0| = \frac{n \cdot (n-1)}{2}$.
Geometric functions are particularly popular to model cooling, whereas logarithmic functions are considered too slow for practical application, although they theoretically converge to a global optimum \cite{talbi2009metaheuristics}.
\par
\begin{figure}[ht]
  \centering
  \begin{minipage}[b]{0.49\linewidth}
    \includegraphics[width=\linewidth]{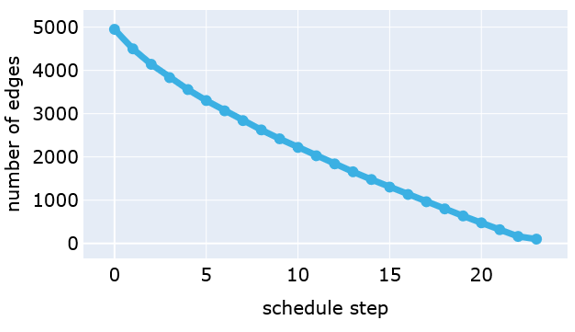}
    \subcaption{$\alpha = 0.1$}
    \label{fig:reduction_sched01}
  \end{minipage}
  \hfill
  \begin{minipage}[b]{0.49\linewidth}
  \includegraphics[width=\linewidth]{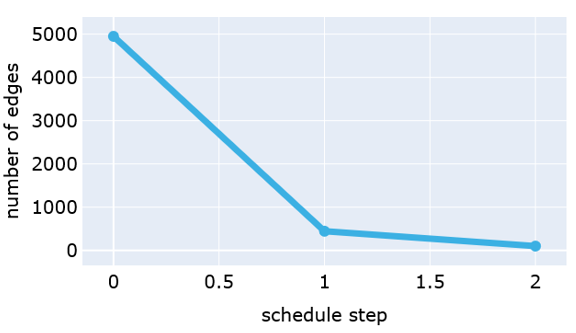}
    \subcaption{$\alpha = 1$}
    \label{fig:reduction_sched1}
  \end{minipage}
  \caption{Reduction schedule with 100 agents and different $\alpha$}
\end{figure}

 Considering that the initial number of edges is much smaller than usual starting temperatures, a slower reduction seems appropriate. A combination of linear, geometric and logarithmic reduction functions was chosen. \autoref{eq:reduction} displays the function that determines the number of edges $\delta$ at each schedule step. 
\begin{equation}
\label{eq:reduction}
  \delta_{i+1} = |E_i| - \frac{|E_0|}{log(i)} \cdot \alpha, \quad with \enspace \alpha \in \; ]0,1]
\end{equation}
The index $i$ represents the index of the step in the reduction schedule. $\delta_i$ determines the number of edges that should remain in the reduction schedule step $i$. 
Thus the new communication topology graph is constructed such that
\begin{equation}
    G_{i+1} = (V, E_{i+1}),  \quad with \quad |E_{i+1}| =  \delta_{i+1}
\end{equation} 
The parameter $\alpha$ controls how many steps the removal schedule includes. If it is close to 0, only a few edges are removed in each step, leading to a slow decrease of connectivity. If it equals 1, the number of edges is reduced in large steps, which leads to a rapid decrease in connectivity. The edges that are removed are selected randomly, ensuring that the final ring topology remains. 
The figures \ref{fig:reduction_sched01} and \ref{fig:reduction_sched1} show how the parameter $\alpha$ influences the granularity of the edge reduction schedule.
\par Finally, the transition criterion from one topology to the next must be specified.
It defines when a transition from one schedule step to the next occurs.
Common conditions for a temperature adaptation in SA are counting of iterations, acceptances, rejections or a combination of these \cite{talbi2009metaheuristics}.
In this case, a simple approach is used, where the number of local searches (equivalent to the number of iterations) since the last transition is counted. As soon as this number becomes larger than $n$, the topology is adjusted. Since the starting topology is a complete graph all agents start at the same time and each unit performs  its local search at least once before the first adjustment. In later stages, this procedure can lead to some agents optimizing locally multiple times between topology adjustments and others not optimizing at all. 
This rapid transition was chosen since it showed the best results in preliminary experiments.
\par To evaluate this dynamic approach, we perform a systematic comparison of the performance of the approach in different parameterizations and several static topologies. For this purpose, we use a set of well-known benchmark functions. However, instead of considering them separately, we perform fitness landscape analysis to match the individual difficulties of the problems with the performance of the topologies and find possible correlations.
Consequently, we first explain the metrics used in the following section before moving on to the experimental setup and the evaluation of the results.

\section{Fitness Landscape Analysis}
\label{sec:landscape}
As the no free lunch theorem states, no optimization heuristic can be superior to all others without regard to the problem \cite{wolpert1997no}.
Different communication topologies significantly influence the information propagation in a heuristic and thus the resulting optimization process. 
Consequently, for different problems, different topologies presumably lead to more advantageous behavior.
We use various fitness landscape metrics to classify the objective functions to examine the relationships between problem characteristics and the performance of different topologies.
\par A fitness landscape, as presented e.g.\ in \cite{sun2014selection}, is defined by the search space $X$, containing all possible solutions of the problem, connected according to a defined distance measure, and the fitness function $f: X \to R$. A fitness landscape for a continuous problem, often uses euclidean distance measures and thus can be described as a landscape with the search space as the bottom floor and the landscape surface being elevated according to the values of the fitness function $f$ \cite{talbi2009metaheuristics}.
Analogous to a geographical landscape, fitness landscapes can have peaks, valleys, plains, canyons, cliffs, plateaus, basins, etc. Investigating these landscape characteristics provides clues as to how difficult it is to find an optimum, i.e., the highest mountain peak (maximization) or the lowest valley (minimization). \autoref{fig:landscapes} shows 3-D plots of benchmark functions demonstrating some manifestations of such landscape features.  
\par Various metrics have been proposed in literature. We limit the scope to metrics that can be used for continuous search spaces and that can be normalized, since we want to be able to compare different functions. In \cite{sun2014selection}, Sun et al. distinguished some basic features of fitness landscapes and argued that for proper characterization, these features must be covered when selecting a set of metrics. These features include:
\begin{itemize}
    \item Dimensionality 
    \item Separability
    \item Ruggedness, Smoothness and Neutrality
    \item Modality 
    \item Deception and Evolvability
\end{itemize}
The dimensionality of the problems is a selectable parameter in the experimental setup and therefore known. 
Furthermore, separability is a well-known property of the benchmark functions.
For the other characteristics mentioned, suitable metrics must be selected. We discuss our choice in the following. 
\subsection{Ruggedness, smoothness and neutrality} The characteristics of ruggedness and smoothness concern the quantity and distribution of the local optima in the search space. The fitness differences in a neighborhood can be large (rugged), small (smooth), or barely present (neutral). Each of these surface shapes presents different challenges for optimization algorithms. 
In \cite{malan2009quantifying} Malan and Engelbrecht adapted the entropy based measure for ruggedness that was first proposed by Vassilev et al. \cite{vassilev2000information} for continuous fitness landscapes. The information theoretic technique is based on a random walk through the search space. 
The random walk is represented as a string with respect to the information stability measure $\epsilon$. If the magnitude of the difference between two fitness values is less than $\epsilon$, they are considered to be equivalent. The string representation is obtained as follows:
\begin{equation}
\label{eq:eps_walk}
    S_i(\epsilon) = \begin{cases}
      -1, & \text{if $f_i - f_{i-1} < - \epsilon$}\\
      0, &  \text{if $|f_i - f_{i-1}| \leq  \epsilon$}\\
      1, &  \text{if $f_i - f_{i-1} > \epsilon$}\\
    \end{cases}       
\end{equation}
The entropy value is calculated for the resulting string $S_i(\epsilon)$ according to \autoref{eq:entropy_rugged}:
\begin{equation}
\label{eq:entropy_rugged}
    H (\epsilon) = - \sum_{p\neq q} P_{[pq]} log_6 P_{[pq]}
\end{equation}
where $P_{[pq]}$ is the frequency of occurrence of the block $[pq]$ in $S_i(\epsilon)$ with $p,q \in \{-1, 0, 1\}$.
This entropy is calculated with various $\epsilon$ between $0$ and $\epsilon_{max}$, which is the value at which the resulting string consists only of zeros. The maximum of all attained entropy measures is taken as the final result \cite{malan2009quantifying}. This entropy measure reflects the information content of the random walk. This is naturally high for a rugged landscape, whereas a smoother landscape has a smaller entropy value.
\par Depending on the stepsize of the random walk, ruggedness can be viewed on different scales. We follow Malan's suggestion and compute the metric once based on random walks with a maximum step size of 1\% of the search space and once with 10\%.
The resulting metrics $FEM_{micro}$ and respectively $FEM_{macro}$ (First Entropic measure as defined by \cite{vassilev2002smoothness} ) are values in $[0,1]$ and reflect the relationship between ruggedness and neutrality on micro an macro scale.
\par Similarly to the $FEM$ a second entropy measure can be calculated that estimates the smoothness of the function rather than the ruggedness. Based on the string $S_i(\epsilon)$ the entropy of smooth blocks, i.e. two consecutive characters with the same sign, is calculated as follows:
\begin{equation}
    h(\epsilon) = - \sum_{p = q} P_{[pq]} log_3 P_{[pq]}
\end{equation}
We apply the same approach as for $FEM$ by calculating $h(\epsilon)$ with different values for $\epsilon$ and keeping the maximum of all entropy values as $SEM$ (Second Entropy Measure).
The $SEM_{micro}$ and $SEM_{macro}$ are again values in $[0,1]$, but refer to the interaction of smoothness and neutrality of the landscape \cite{vassilev2002smoothness}.
\subsection{Modality}
The \textit{modality} of a function corresponds to the number of local optima. Modality and ruggedness are closely related, since a rugged landscape may also include many local optima. In  \cite{vassilev2000information}, Vassilev et al. also proposed a metric to quantify the modality of the random walk encoded by \autoref{eq:eps_walk}. A new string $S'(\epsilon)$ is constructed by removing all zeros from $S(\epsilon)$ and reducing sequences of equal characters to one character. Thus, $S'(\epsilon)$ contains only information that is essential with respect to modality. The resulting modality measure is called partial information content (PIC) and is given by
\begin{equation}
    PIC(\epsilon) = \frac{\mu}{n}
\end{equation}
where $n$ is the length of $S(\epsilon)$ and $\mu$ the length of $S'(\epsilon)$.
If the random walk encountered a landscape with high modality, the length of $S'(\epsilon)$ is almost the same as that of $S(\epsilon)$, resulting in $PIC(\epsilon)$ being close to or equal to one. If the path is flat or only leading in one direction $PIC(\epsilon)$ tends to or is equal to zero. 
For $PIC$, we use the same procedure as for $FEM$ and $SEM$ and perform random walks with varying step sizes to look at modality at different scales. Accordingly, we use $PIC_{micro}$ and $PIC_{macro}$ as metrics. 
\subsection{Deception and evolvability}
A \textit{deceptive} landscape provides information that can guide an optimization algorithm away from the global optimum, towards local optima. Which properties of a function are deceptive depends on the algorithm - in the case of distributed algorithms, perhaps also on the communication topology. 
One possibly deceiving characteristic is the presence of funnels.
A funnel is a cluster of local optima that forms a global basin shape \cite{malan2013ruggedness}. 
The dispersion metric of Lunacek et al. \cite{lunacek2006dispersion} provides insight into the global topology of fitness functions and thus indirectly allows estimation of the presence of funnels. 
In \cite{malan2013ruggedness}, Malan and Engelbrecht proposed a normalized version to allow comparison of functions with different domain sizes.
To compute the dispersion metric, a random sample $\mathscr{S}$ of length $n$ is drawn that is uniformly distributed over the search space. From this sample $\mathscr{S}$, a subset $\mathscr{S}^*$ is determined that contains the $\mathscr{s}$ best points by fitness values. To make functions with different domain sizes comparable, the position vectors of $\mathscr{S}^*$ are normalized in such a way that the search space is scaled to [0,1]. 
In addition, a comparison sample $\mathscr{C}$ also of size $\mathscr{s}$ is sampled uniformly across the search space.
Let $disp(\mathscr{S})$ be the average pairwise distance between normalized positions in the sample $\mathscr{S}$. Then the dispersion metric $DM$ is defined as follows:
\begin{equation}
    DM = disp(\mathscr{S}^*) - disp(\mathscr{C})
\end{equation}
Thus, the metric quantifies how far points with high fitness values are away from each other compared to a large uniform random sample.
It yields values in the range of [-1,1]. A low value ($DM < 0$) indicates a single funnel landscape with an underlying unimodal structure. A high value ($DM > 0$) indicates a multi-funnel landscape and underlying multimodal structure.
\begin{figure}[htbp]
\caption{Selection of benchmark functions as 3-D plot on the full domain (fd) or on a 1 \% section of the domain (1\%d)}
\label{fig:landscapes}
  \centering
  \begin{subfigure}[b]{0.32\linewidth}
    \includegraphics[width=\textwidth]{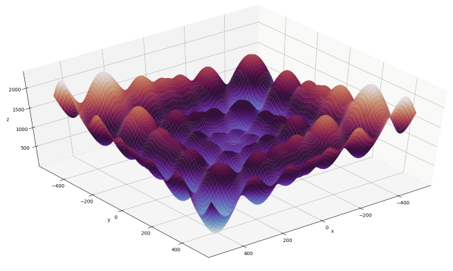}
    \caption{Schwefel 2.26 with penalty (fd)}
    \label{fig:schwefel_pen_fd}
  \end{subfigure}
  \hfill
  \begin{subfigure}[b]{0.32\linewidth}
    \includegraphics[width=\textwidth]{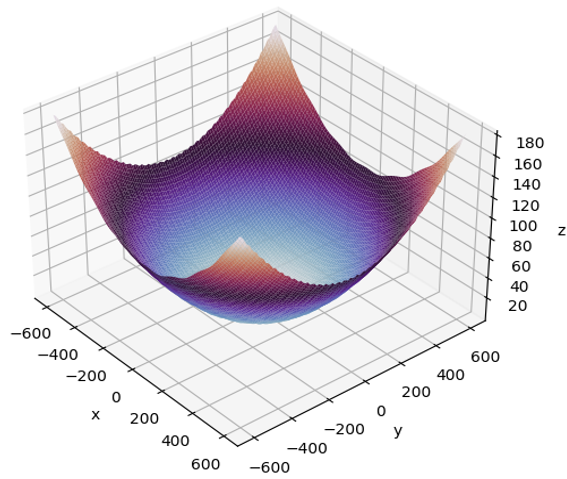}
    \caption{Griewank (fd)}
    \label{fig:griewank_df}
  \end{subfigure}
    \hfill
   \begin{subfigure}[b]{0.32\linewidth}
    \includegraphics[width=\textwidth]{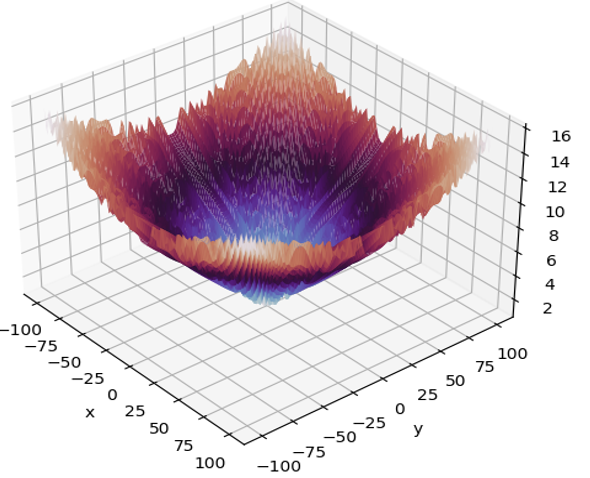}
    \caption{Salomon (fd)}
    \label{fig:salomon_fd}
  \end{subfigure}
\hfill
     \begin{subfigure}[b]{0.32\linewidth}
    \includegraphics[width=\textwidth]{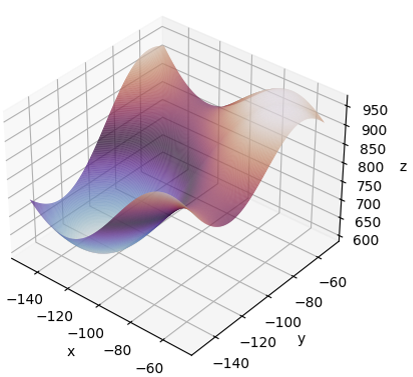}
    \caption{Schwefel 2.26 with penalty (1\%d)}
    \label{fig:schwefel_pen_1d}
  \end{subfigure}
    \hfill
     \begin{subfigure}[b]{0.32\linewidth}
    \includegraphics[width=\textwidth]{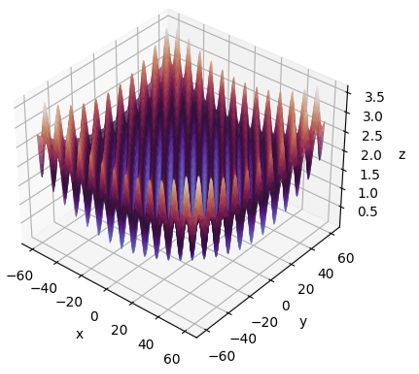}
    \caption{Griewank (1\%d)}
    \label{fig:griewank_1d}
  \end{subfigure}
  \hfill
  \begin{subfigure}[b]{0.32\linewidth}
    \includegraphics[width=\textwidth]{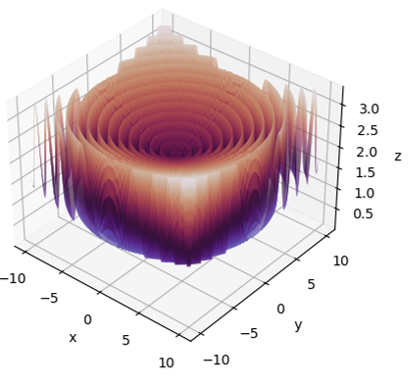}
    \caption{Salomon (1\%d)}
    \label{fig:salomon_1d}
  \end{subfigure}
\end{figure}
\autoref{fig:landscapes} shows 3-d plots of three benchmark functions once in their full domain and once in a one-percent section of their domain, respectively.
The penalized Schwefel 2.26 function is very rugged on the macro scale (\autoref{fig:schwefel_pen_fd}), but much less so on the micro scale (\autoref{fig:schwefel_pen_1d}). The function has a high value in the dispersion metric, which is also consistent with its global multi funnel shape. In contrast, both Griewank and Salomon have a small value in the dispersion metric, corresponding to their global single funnel shapes. While Salomon is highly rugged on macro and micro scale with a slight increase on micro scale, Griewank is much more rugged on micro scale than on the macro level. This effect actually decreases in higher dimensions, making the function "easier" to solve \cite{locatelli2003note}. 

\autoref{tab:metric_overview} shows an overview of the applied fitness landscape metrics, including a brief summary. The results obtained for the metrics for the benchmark functions are summarized in a dedicated repository\footnote{\url{https://github.com/sholly-offis/Deta}}.
\begin{table}[htbp]
    \centering
    \renewcommand{\arraystretch}{1.3}
    \begin{tabularx}{\linewidth}{ |c|X| }
        \hline
        dimension & dimension, here also equal to number of agents\\
        \hline
        separability & boolean that indicates if variables of a function are independent \\
        \hline
        $FEM_{macro}$ & \multirow{2}{\hsize}{first entropy based measure of \textit{ruggedness} on macro and micro scale} \\
        $FEM_{micro}$ &  \\
        \hline
        $SEM_{macro}$ & \multirow{2}{\hsize}{second entropy based measure of \textit{smoothness} on macro and micro scale} \\
        $SEM_{micro}$ &  \\
        \hline
        $PIC_{micro}$ & \multirow{2}{\hsize}{measure of partial information content concerning \textit{modality} on macro and micro scale} \\
        $ PIC_{macro}$ &  \\
        \hline
        $DM$ & dispersion metric that quantifies distances between good solutions and thus indicates the presence of funnels \\
        \hline
    \end{tabularx}
    \caption{Overview of applied metrics}
    \label{tab:metric_overview}
\end{table}

\section{Methodology}
\label{sec:methods}
The goal of the experimental study is to investigate if a dynamic topology adaptation approach outperforms static topologies.
We use multiple benchmark functions, all scalable and multi-modal, and the distributed optimization heuristic presented in \autoref{subsec:cohda}.
\par Furthermore, we examine correlations between the properties of the benchmark functions and the performance of different topologies in the defined performance dimensions. In doing so, we also consider different parameterizations of the dynamic approach.
Thereby, we hope to find clues that will help in the further development of the dynamic approach and ultimately lead to the parameterization of dynamic topology adaptation in such a way that it provides a tailored solution to a problem. 
In the following we first give a short introduction to the chosen example heuristic and then elaborate on the experimental setup.
\subsection{Distributed Optimization Algorithm}
\label{subsec:cohda}
COHDA is a combinatorial optimization heuristic for distributed agents, and was developed for the self-organized scheduling of distributed energy resources in virtual power plants \cite{hinrichs2017distributed}.
The heuristic can be classified as a system realizing a gossiping protocol based on strictly defined communication and knowledge integration rules. 
In \cite{bremer2017agent}, Bremer et al. adapted COHDA to ﬁnd the global minimum of a real valued objective function.
An agent $a_i$ is responsible for only one value $x_i$ from a continuous search space. It performs its local optimization to minimize the global objective function, by adapting its own choice of $x_i$ while considering the choices of other agents $x_j, j \neq i$ as temporarily fixed.
Agents send update messages to their neighbors - as defined by the communication topology - to pass on new information from their neighbors or to inform about their own changes in the value selection. 
Depending on well-defined convergence conditions, COHDA has been proven to always converge at least to a local optimum \cite{hinrichs2017distributed}. Parts of these conditions are related to the topology, and thus have to be reflected here: The chosen topology has to be connected, irreflexive, and symmetric. As a consequence, the topology adaptation developed in this work has to guarantee these characteristics in all intermediate stages to not sacrifice convergence.
\subsection{Experimental Setup}
The experiments are preformed with agent systems in two different sizes, namely 50 and 100. In the applied setup, the system size is equal to the dimension of the objective functions, as each agent is responsible for choosing one solution variable.
A set of 13 different benchmark functions is used as underlying synthetical problem instances. They include Ackley, a scalable version of Eggholder, Griewank, Happy Cat \cite{beyer2012happycat}, Rana, Rastrigin \cite{li2013benchmark}, Rosenbrock, Salomon, Schaffer F6, Qing, Schwefel 2.26 and a penalized Version of Schwefel 2.2.6 which was introduced in \cite{holly2020effects}. Unless otherwise stated, the definitions are taken from \cite{momin2013literature}. Definitions, domains and global minima of the benchmark functions are listed in a dedicated repository\footnote{\url{https://github.com/sholly-offis/Deta}}. 
\par
\begin{figure}
\caption{Overview of reflected static communication topologies}
\label{fig:topos}
  \centering
  \begin{subfigure}[b]{0.32\linewidth}
    \includegraphics[width=0.8\textwidth, height=5em]{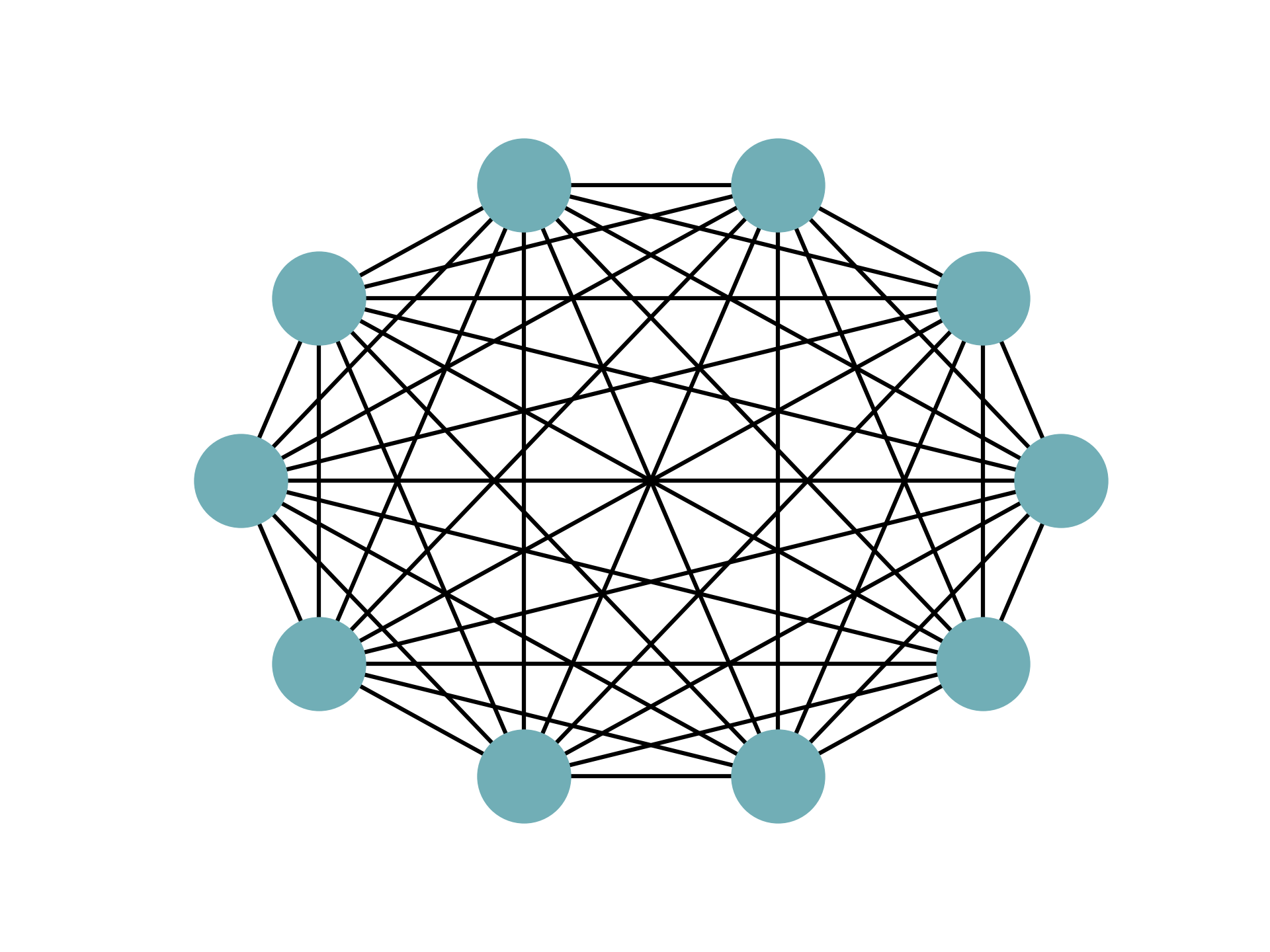}
    \caption{complete graph}
    \label{fig:complete}
  \end{subfigure}
  \hfill
  \begin{subfigure}[b]{0.32\linewidth}
    \includegraphics[width=0.8\textwidth, height=5em]{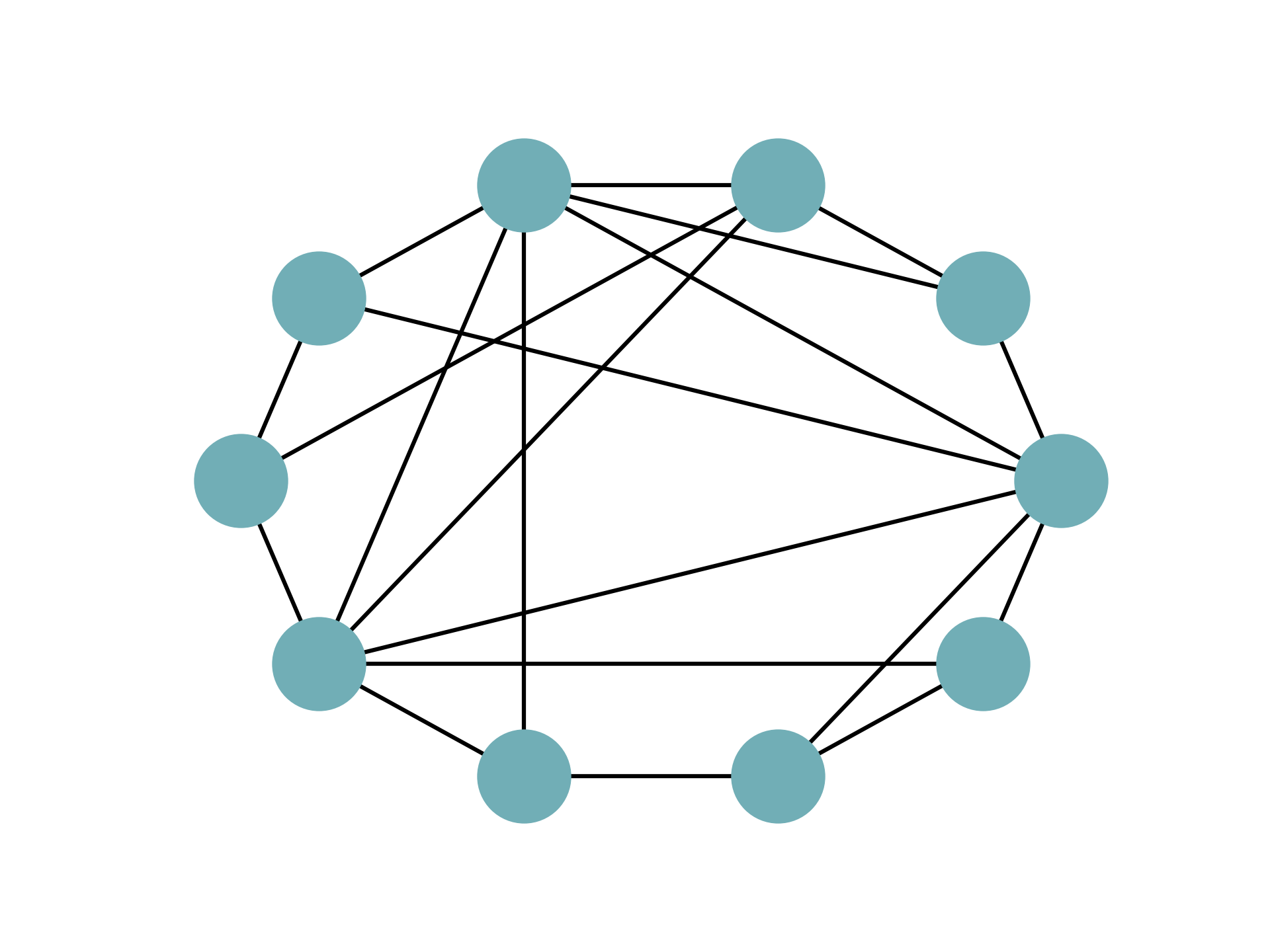}
    \caption{small world topology}
    \label{fig:smallworld}
  \end{subfigure}
    \hfill
  \begin{subfigure}[b]{0.32\linewidth}
    \includegraphics[width=0.8\textwidth, height=5em]{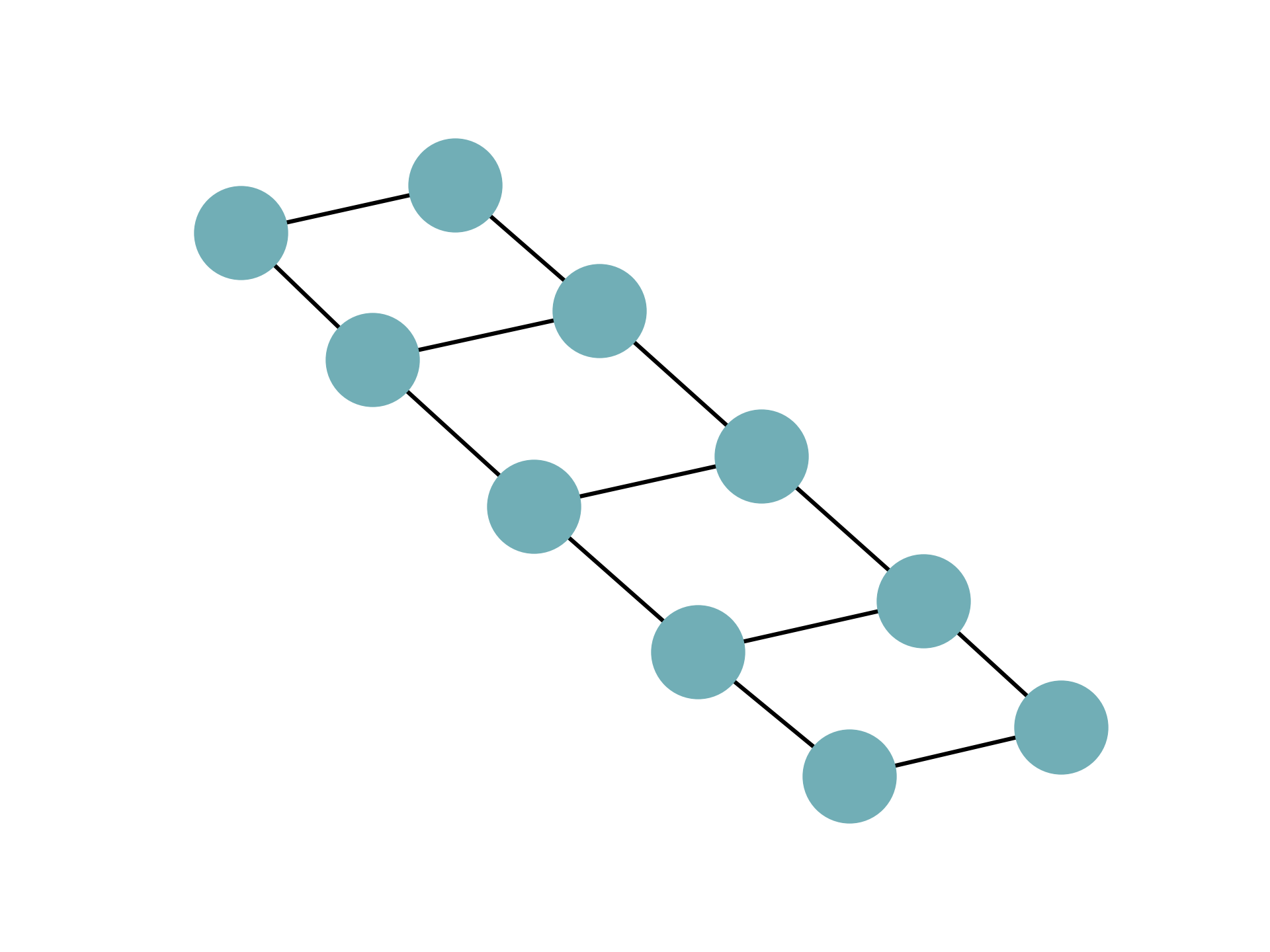}
    \caption{grid}
    \label{fig:grid}
  \end{subfigure}
   \centering
  \begin{subfigure}[b]{0.32\linewidth}
    \includegraphics[width=0.8\textwidth, height=5em]{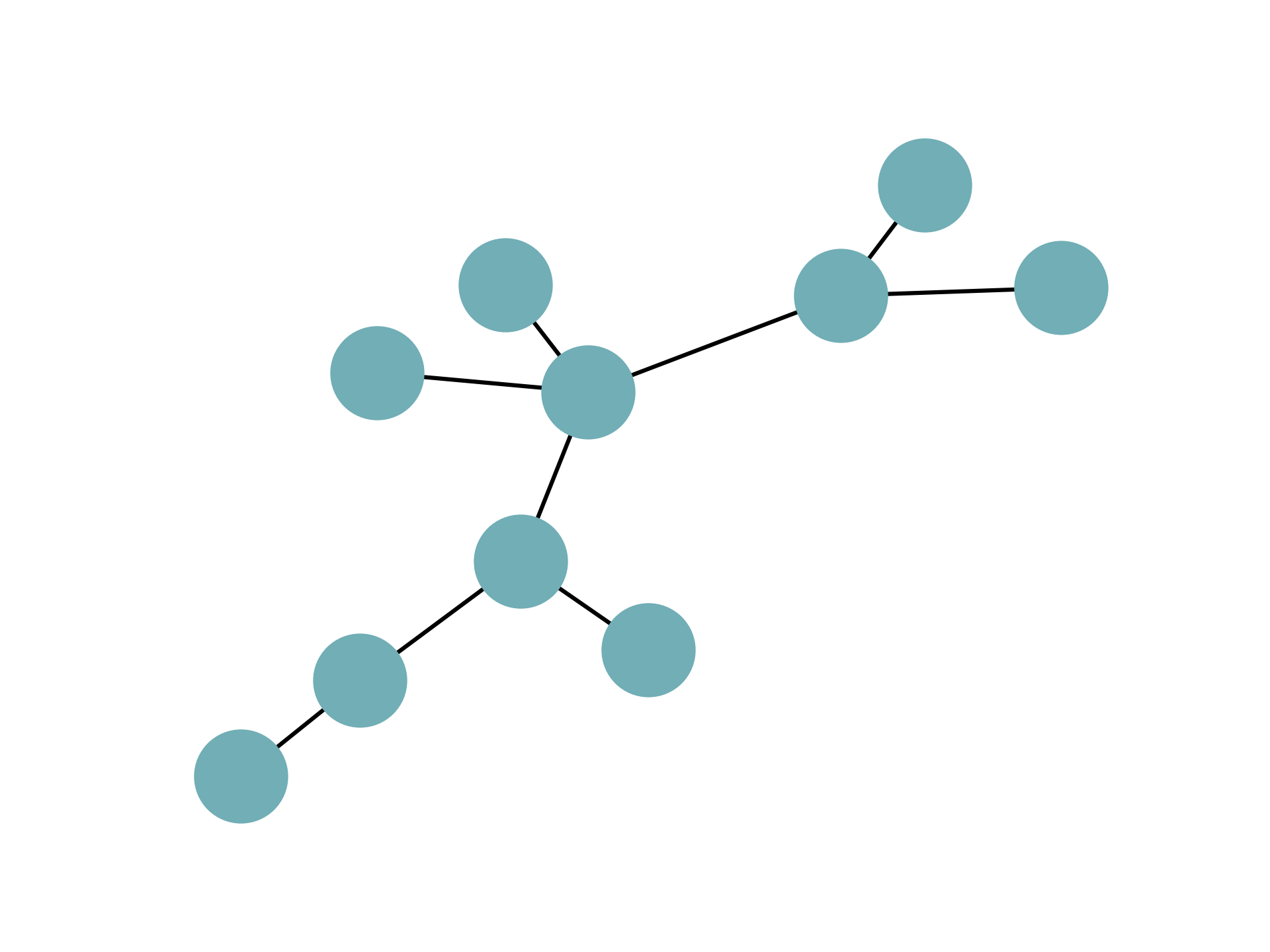}
    \caption{random tree}
    \label{fig:tree}
  \end{subfigure}
  \hfill
  \begin{subfigure}[b]{0.32\linewidth}
    \includegraphics[width=0.8\textwidth, height=5em]{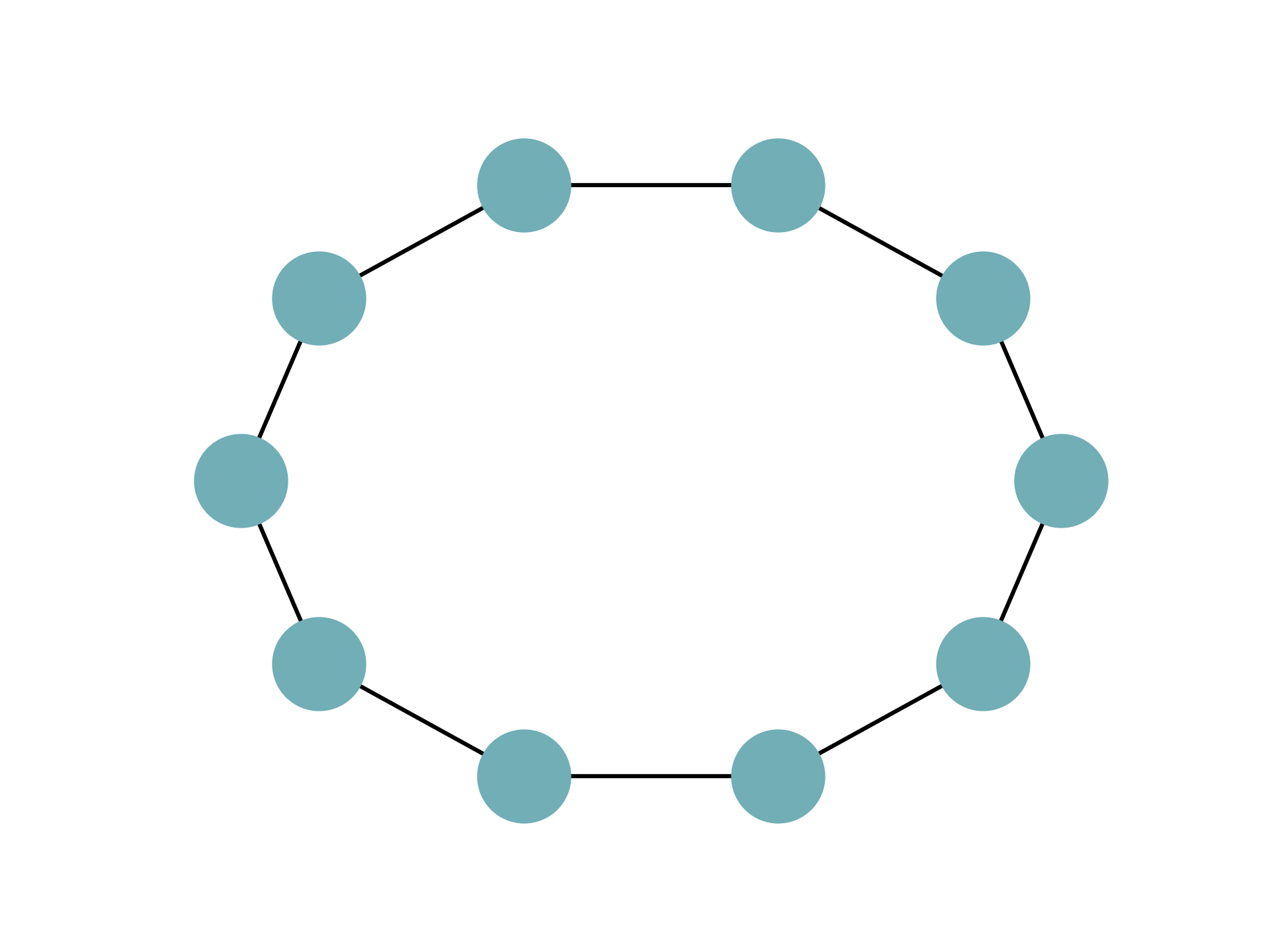}
    \caption{ring}
    \label{fig:cycle}
  \end{subfigure}
    \hfill
  \begin{subfigure}[b]{0.32\linewidth}
    \includegraphics[width=0.8\textwidth, height=5em]{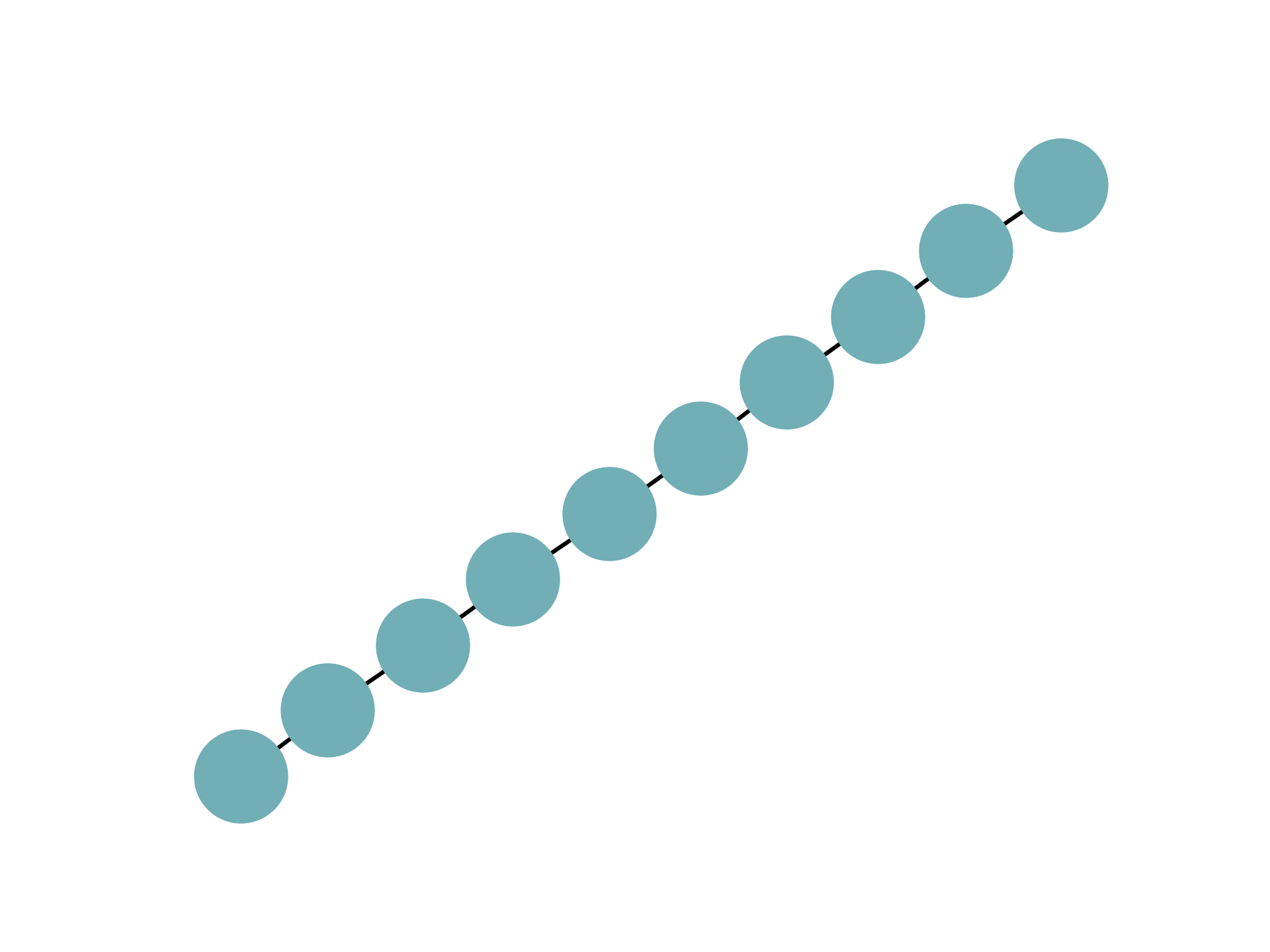}
    \caption{path graph}
    \label{fig:path}
  \end{subfigure}
   \vspace{-1.5em}
\end{figure}

 The search process in the solution space defined by each benchmark function is performed using COHDA and different topologies.
This involves complete graphs, ring-, tree-, small-world-, path-, and grid- topologies. 
\autoref{fig:topos} shows these topology types for 10 agents.
\par For each system size (50 and 100) each topology type is created, including a randomized setting with 5 different seeds and optimization runs and 10 different starting seeds.
As a result of this setting, 50 different setups are examined for each combination of the number of agents and topology type. 
The same is done for the dynamic approach with different values for $\alpha \in [0.1, 0.3, 0.7, 1]$. \footnote{$\alpha = 0.5$ was discarded in this presentation for reasons of brevity as it was not superior to the other values.}
A total of 1000 optimization runs were performed per benchmark function (600 with static topologies and 400 with the dynamic approach). Therefore, the following evaluations are based on a total of 13,000 optimization runs. 
\par To compare the performance, four performance dimensions are evaluated:
\begin{itemize}
    \item solution quality: error measure
    \item speed of convergence: time required to converge
    \item communication traffic: number of messages exchanged between agents
    \item computational effort: number of local searches preformed by all agents
\end{itemize}
\par Each of the performance dimensions is normalized per system size and benchmark function. We differentiate the four performance dimensions when determining the best topology for a benchmark function. 
To be the best topology in a performance dimension, a topology must achieve the lowest possible values for this measure. Therefore, we consider the mean in each case. 
For some topologies however, the spread of values is extremely large, so that despite a good mean value, the risk of an unfavorable outlier is much higher than for other topologies. Again, in the application domain of critical infrastructure optimization, this is unfavorable.
The sum of mean and standard deviation is used as an additional measure (using the notion \textit{mps} -- mean plus standard deviation), to reflect this aspect.
\par To relate the performance of the topologies to the characteristics of the benchmark problems, fitness landscape metrics described in \autoref{sec:landscape} are computed for each benchmark function using random walks of length 1000 and the mean of 30 runs is taken as measure as proposed by \cite{malan2013ruggedness}.
Finally, we train decision trees using the CART algorithm \cite{breiman1984classification} implemented by scikit-learn \cite{scikit_learn2011}. We use them to determine which metrics can be employed to distinguish benchmark functions and assign them to the best-performing topologies.
\section{Results and Discussion}
\label{sec:results}
In the following, we first examine the differences in the solution quality, using decision trees generated on the basis of the fitness landscape analysis. 
Additionally, we want to identify correlations between the properties of the fitness landscapes and the performance of different topologies and parameterizations of the dynamic adaptation approach.
\par In the second part of the evaluation, we examine the other performance dimensions. Since these cannot be reasonably analyzed on their own, individual functions are analyzed as representatives in order to investigate the overall performance of the different topologies and the trade-offs between the performance dimensions. 
\subsection{Decision tree-based performance analysis}
To determine which topology is best for a given function and dimension for a performance indicator, minimum mean and mps are evaluated in each case. For simplicity, additional collective topology categories were introduced when several topologies were equally good. All topologies whose mean and mps do not deviate more than 1\% from the minimum values are assigned to a best list. 
If the best list contains more than three entries, single topologies are replaced by collective categories. 
The label \textit{highly meshed} is assigned if only highly meshed topologies are in the given class (complete, small world, grid, dynamic). \textit{Weakly meshed} accordingly summarizes ring, tree, and path graph topologies. Other combinations are labeled as \textit{various}. Note that if $dynamic$ is displayed without a value for $\alpha$, multiple values for $\alpha$ have performed equally well. 
In order to classify what a decision boundary for a particular landscape metric means, the distribution of the metric in the set of benchmark functions must be considered. For this purpose, \autoref{tab:dist_metrics} shows the respective minima, maxima, mean values and the limits for the upper and lower quartiles. 
\begin{table}[ht]
    \centering
      \renewcommand{\arraystretch}{1.2}
  \begin{tabular}{ |c|c|c|c|c|c| }
\hline
 & min & $Q_1$ & median/$Q_2$ & $Q_3$ & max\\
\hline
$DM$ & -0.39 & -0.28 & -0.2 & -0.01 & 0.05\\
\hline
$FEM_{macro}$ & 0.69 & 0.72 & 0.87 & 0.88 & 0.89\\
\hline
$FEM_{micro}$ & 0.2 & 0.4 & 0.66 & 0.76 & 0.9\\
\hline
$SEM_{macro}$ & 0.51 & 0.53 & 0.57 & 0.68 & 0.74\\
\hline
$SEM_{micro}$ & 0.52 & 0.65 & 0.73 & 0.78 & 0.8\\
\hline
$PIC_{macro}$ & 0.35 & 0.38 & 0.61 & 0.66 & 0.7\\
\hline
$PIC_{micro}$ & 0.07 & 0.14 & 0.32 & 0.46 & 0.74\\
\hline
\end{tabular}
    \caption{Distribution of fitness landscape metrics}
    \label{tab:dist_metrics}
\end{table}
\begin{figure*}[ht]
  \centering
  \includegraphics[width=0.9\linewidth, trim={3em 3em 3em 3em}]{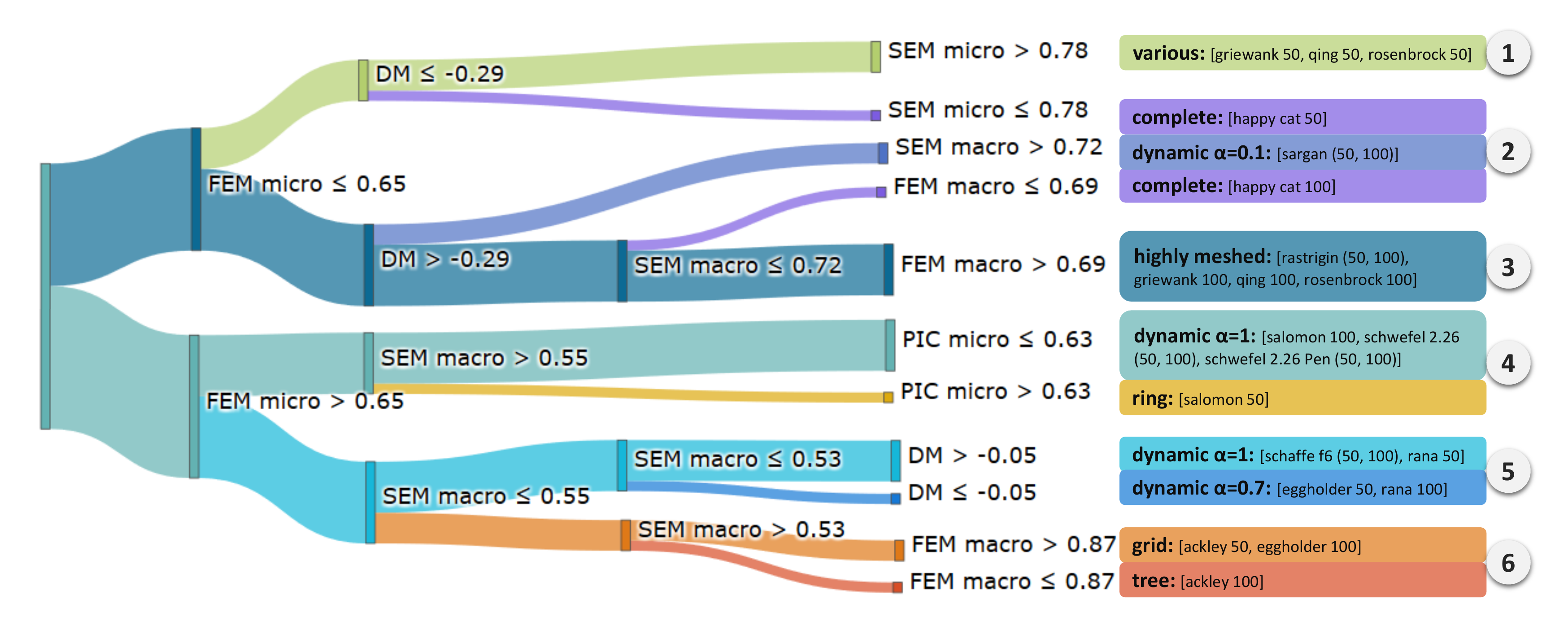}
  \caption{Decision tree for mean error}
  \label{fig:tree_error_mean}
\end{figure*}
\begin{figure*}
    \centering
    \includegraphics[width=0.9\linewidth,  trim={3em 3em 3em 3em}]{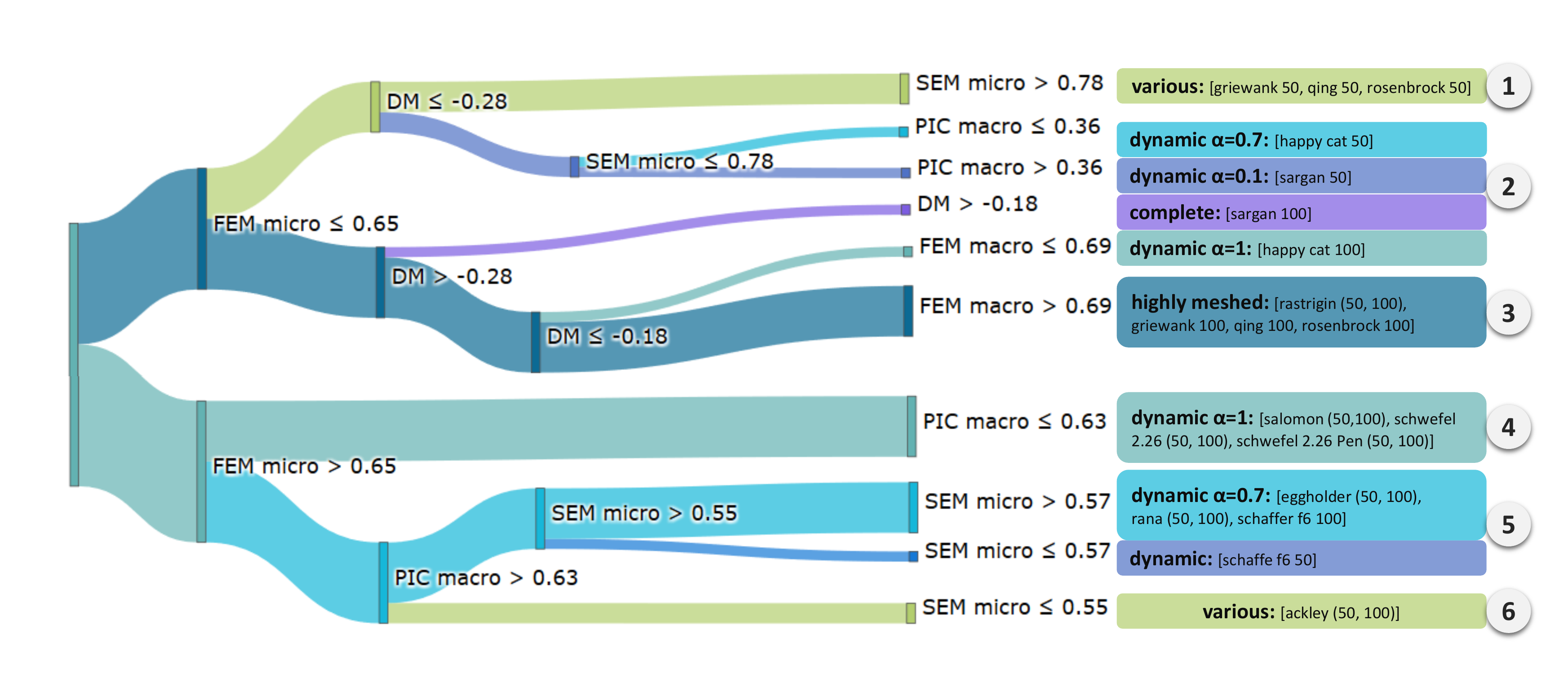}
    \caption{Decision tree for mps error}
    \label{fig:tree_error_mps}
\end{figure*}
\par First, we focus on the achieved solution quality in terms of the normalized error.
\autoref{fig:tree_error_mean} shows the resulting decision tree as sankey diagram for the mean error and \autoref{fig:tree_error_mps} the tree illustrating the error mps.
The thickness of the branches indicates the number of benchmark functions that belong to a path. On the right, the best topologies or topology categories are shown with the functions they contain. To facilitate the discussion of the results, individual end nodes or clusters of end nodes were numbered. These are discussed below.
\begin{enumerate}
\item functions with below average micro ruggedness ($FEM_{micro}$) and small dispersion ($DM$), i.e. good points in the search space are close to each other; These function are easy to solve, thus many topologies provide good results%
\item Sargan and Happy Cat function; The Sargan function has the highest values for macro smoothness ($SEM_{macro}$). The Happy Cat function is also very smooth. For both functions complete graph and the dynamic approach perform similarly, while all other topologies rank far behind.%
\item functions with below average micro ruggedness but higher dispersion; slightly more difficult than (1), thus weakly meshed topologies are outperformed.%
\item functions with above average micro ruggedness ($FEM_{micro}$), above average macro smoothness ($SEM_{macro}$) and low to medium modality on macro scale ($PIC_{macro}$) are solved best by $dynamic_{\alpha=1}$ topology. Solomon (50) has a very high micro ruggedness. If only the mean error is considered, the ring topology performs best. However, if the reliability is also considered (see \autoref{fig:tree_error_mps}), the $dynamic_{\alpha=1}$ topology performs better. The ring topology is obviously very well suited for the intensive exploitation of landscape areas with high micro ruggedness and lower macro smoothness, but an initial exploration push, as in the dynamic approach, helps to guide the search more reliably into good regions.
\item functions with above average micro ruggedness and macro smoothness in the lower quartile; dynamic topology adaptation with different values for $\alpha$ achieves the best results there;%
\item functions with above average micro ruggedness but very low smoothness on micro scale but not on macro scale; Considering only the mean error, the grid and tree topologies perform best, but their advantage is very small, so for the mps value for the Ackley function, many topologies perform equally well and Eggholder (100) is best and most reliably optimized with the dynamic topology with $\alpha = 0.7$.%
\end{enumerate}
\par Note that functions for which $dynamic_{\alpha=1}$ topology performs best, have the highest values for the dispersion metric. A $DM$ larger than $-0.037$ seems to be a good indicator for a class of functions with multi-funnel shapes for which the dynamic approach with large values for $\alpha$ achieves the best results. 
Fitness landscapes with high dispersion, high modality, high ruggedness and low smoothness less strongly meshed topologies are advantageous, which often makes the dynamic approach with large values for $\alpha$ favorable. 
For landscapes with lower modality, less dispersion, a less rugged surface, and more smooth sections, more strongly meshed topologies tend to have an advantage, and the dynamic approach with small values for $\alpha$ also often performs well. More analysis on this will be needed in future work. 
\subsection{Examination of further performance dimensions}
 For the analysis of the other performance dimensions, an isolated consideration of the decision trees is not suitable, since there are usually correlations between good and poor performance in several dimensions. For example, the heuristic may have terminated after a short time, but converged to a local optimum. Or only very few messages are sent, but only small parts of the search space are explored, which makes the solution quality a matter of luck. Therefore, we examine the results for the penalized Schwefel 2.26 function, Rana, and Griewank in more detail below.
 \begin{figure}[htbp]
  \centering
  \includegraphics[width=0.8\linewidth]{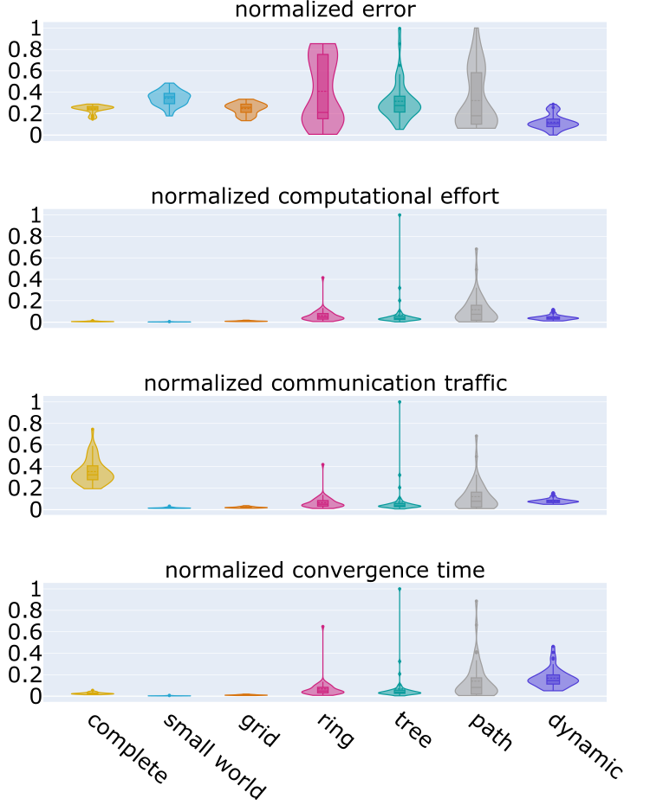}
  \caption{Results for penalized Schwefel 2.26 (100 agents, $dynamic_{\alpha=1}$)}
  \label{fig:vio_schwefel_pen}
   \vspace{-1.5em}
\end{figure}
\par
\autoref{fig:vio_schwefel_pen} shows the results obtained for Schwefel 2.26 with penalty for a system size of 100 agents. Each of the four performance dimensions considered is presented in a separate row. For each static topology and the dynamic approach with $\alpha=1$ a violin plot is displayed, which shows the distribution with an internal box plot surrounded by a density plot. The top row shows the distributions of the normalized errors. For this function, the dynamic approach performs with the highest solution quality in terms of mean value and outliers. 
Compared to the strongly meshed topologies (complete, small world and grid graph), the computational effort depicted in row two of \autoref{fig:vio_schwefel_pen} is slightly increased. 
This seems reasonable, since the topology converges to a ring topology after a short time and therefore performs more computations (with the design objective of these computations being in a promising region of the search space). A similar picture is obtained for the emerging communication traffic, row three in \autoref{fig:vio_schwefel_pen}. However, the complete graph generates much more communication effort.
In terms of convergence time, the dynamic approach shows slower convergence compared to the topologies that are consistently strongly meshed.
\par
\begin{figure}
\caption{Performance results for Rana and Griewank (50 agents, $dynamic_{\alpha=1}$)}
\label{fig:vio_fig12}
  \centering
  \begin{subfigure}[b]{0.49\linewidth}
   \includegraphics[width=\linewidth]{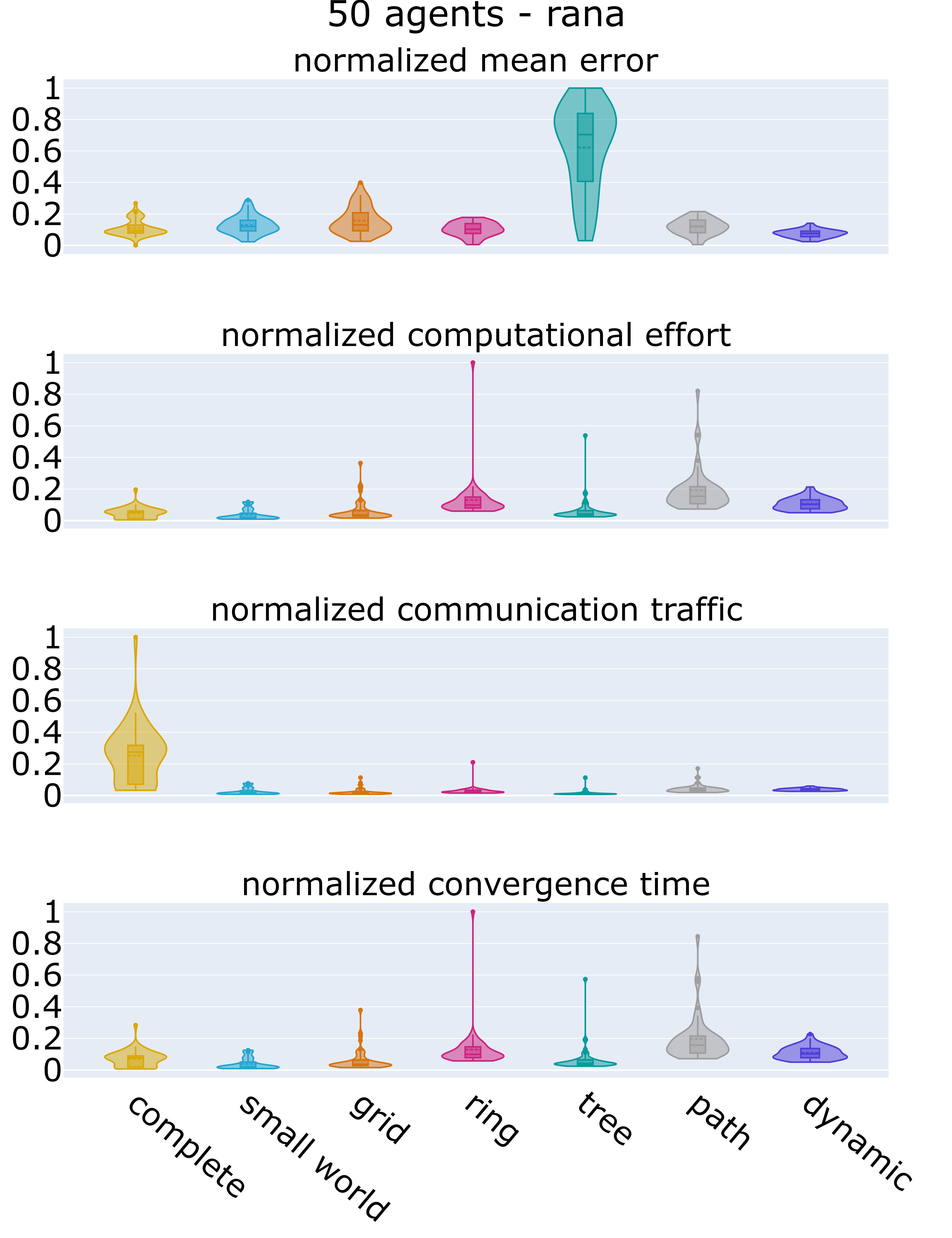}
  \caption{Rana}
  \label{fig:vio_rana}
  \end{subfigure}
  \hfill
  \begin{subfigure}[b]{0.49\linewidth}
   \includegraphics[width=\linewidth]{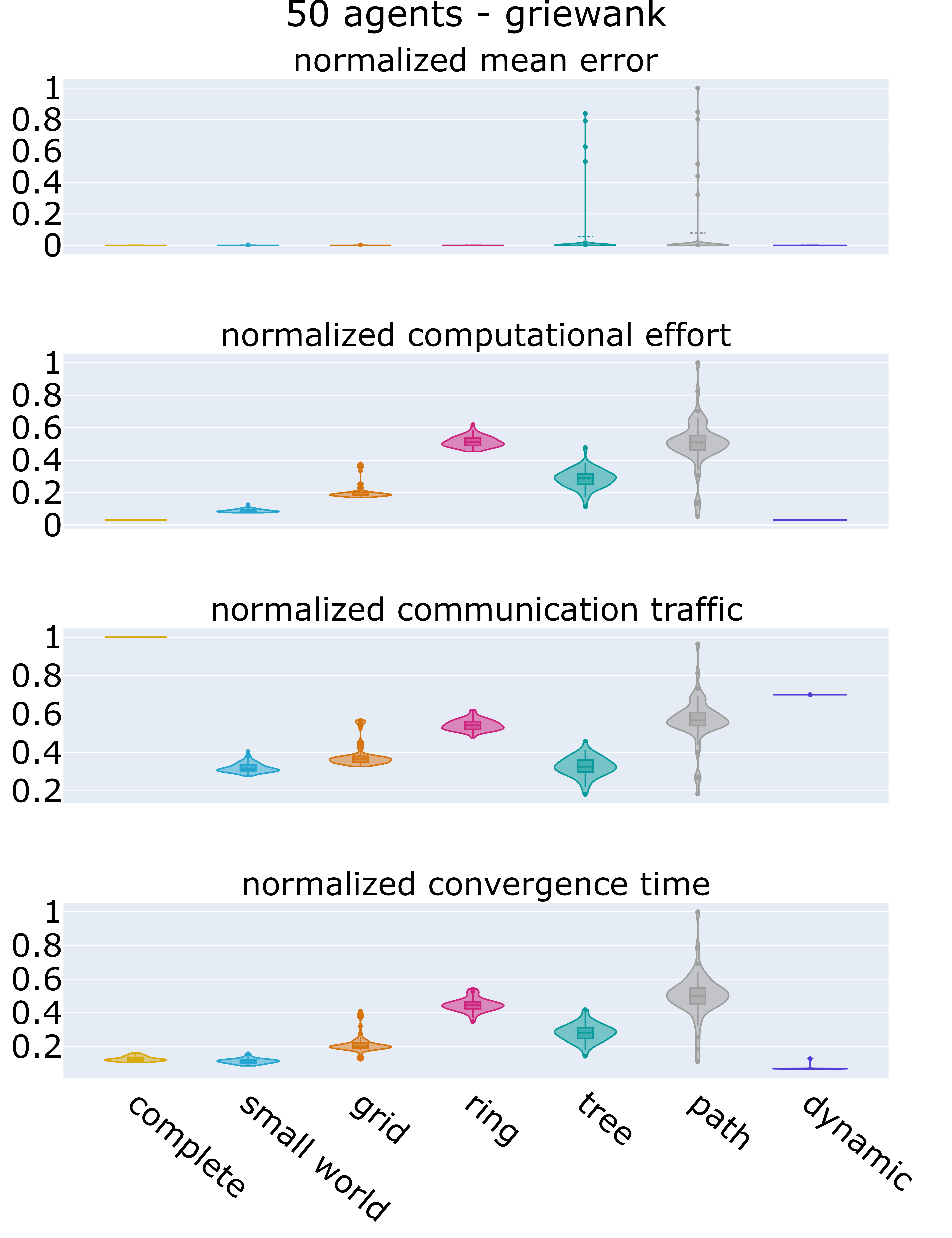}
  \caption{Griewank}
  \label{fig:vio_griewank}
  \end{subfigure}
 \vspace{-1.5em}
\end{figure}

 For Rana, the overall impression is similar to Schwefel 2.26 with penalty.
 The dynamic approach usually performs best and thus needs more time, computational effort and message exchange than other strongly meshed topologies. If the main concern is convergence speed or low communication traffic, the small-word or grid topology are preferable as they provide a good trade-off.
For functions with lower modality, dispersion, and ruggedness, such as Rastrigin or Griewank, where many topologies succeed in finding the global optimum, the complete and dynamic graphs usually converge the fastest and have the lowest computational cost. However, they also generate the largest amount of messages, with the dynamic approach performing slightly better. \autoref{fig:vio_griewank} shows these relationships exemplary for the Griewank function. 
\par To sum up the presented results: In our evaluations, easy to explore fitness landscapes, i.e. functions with uni-modal single funnel shapes and a hardly rugged surface, an arbitrary fairly meshed topology leads reliably to the global optimum. Depending on whether lower message volume or lower computational costs are preferred, small world or the dynamic approach with large $\alpha$ are sufficient. 
For fitness landscapes with multi-modal and multi-funnel shapes, the dynamic approach with a fast reduction schedule shows advantages regarding solution quality with moderate resource consumption.   

\section{Conclusion and Outlook}
\label{sec:outlook}
Distributed optimization heuristics are a suitable approach to handle the increased complexity in cyber-physical multi-energy systems.
The communication between the distributed entities must be carefully designed to ensure reliable behavior of a heuristic and thus its suitability for applications in critical infrastructures like energy systems.
The communication topology defines which entities exchange information and therefore affects solution quality, convergence speed and collaboration costs.
A dynamic approach to topology adaptation during runtime has been presented, based on the principles of simulated annealing. This approach was evaluated against several static topologies using a distributed optimization heuristic to optimize a set of well-known benchmark functions. 
In addition, we conducted a fitness landscape analysis and trained decision trees to derive correlations between problem properties and performance of the communication topologies.
\par The main findings can be summarized as follows:
\begin{itemize}
    \item  Functions with small ruggedness at micro level can be considered as "simple". Moreover, if they have a very small $DM$, i.e., a unimodal shape, many topologies will find the global optimum. Otherwise, more meshed topologies are advantageous.
    \item When low micro ruggedness is combined with high smoothness or very low macro ruggedness, strongly meshed topologies such as the full graph or the dynamic approach with small values for $\alpha$, i.e., slowly decreasing connectivity, outperform other topologies.
    \item The dynamic approach with large values for $\alpha$ excels in providing superior solution quality for functions with high dispersion and thus a difficult multi-funnel landscape. With respect to the cost of cooperation, the dynamic approach is competitive for both easy and hard to explore problems.
\end{itemize}
Regarding the application domain of distributed control in cyber-physical energy systems, search spaces will have to be analyzed in detail. Given the complexity and non-linearity of the resulting system of systems though, high modality and rugged surfaces seem to be characteristic for many use cases with distributed energy resources, controllable loads and storage systems \cite{niesse2017local}. So far, communication topologies for distributed heuristics have been selected mostly independently from solution space characteristics. Our work is a first step towards a more systematical selection and dynamic adaptation of communication topologies in this regard.
\par In future work, we plan to further enhance the dynamic approach. One aspect of this is a intelligent selection of the connections to be removed, by analysing the graph-theoretical characteristics of the intermediate topologies.
Furthermore, non-monotonic reduction schedules could be beneficial for some problems, i.e., schedules in which the number of edges can also increase again under certain conditions. The starting topology may be varied as well, since starting with complete graphs is not always advantageous. Overall, with all these enhancements, we aim to be able to optimally select these parameters in the future depending on the problem characteristics and prioritization of the performance dimensions. The resulting parametrizable dynamic topology adaptation will be evaluated on real world problems in the energy domain. 
\balance
\section*{Acknowledgment}
We would like to express our sincere gratitude to our college Martin Tröschel, who provided us with advice and support, especially in all data science aspects.
\bibliographystyle{IEEEtran}
\bibliography{literature}
\section*{Appendix}
The Appendix includes an overview of the used benchmark functions in \autoref{tab:benchmark_functions} and the values determined for the applied fitness landscape metrics for theses benchmark functions in \autoref{tab:func_metrics}.

\begin{table*}[ht]
    \centering
     \renewcommand{\arraystretch}{1.2}
    \begin{tabularx}{\textwidth}{|c|>{\setlength{\baselineskip}{2\baselineskip}}X | c |}
         \hline
      \textbf{function} & \textbf{Definition, Domain and global optimum $f(\vec{x}^*)$} & \textbf{3-D Plot} \\ \hline
       \multirow{3}{*}{Ackley}  & $ f(\vec{x}) = - 20 exp(-0.2\sqrt{\frac{1}{n}\sum_{i=1}^{n}x_i^2})-exp(\frac{1}{n}\sum_{i=1}^{n}cos(2\pi x_i))+ a + exp(1)$ & \multirow{3}{*}{ \includegraphics[height=4.1em]{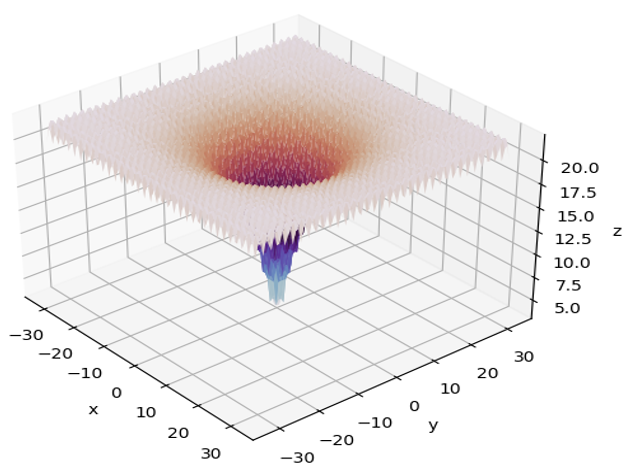}}\\
        & $x_i \in [-32, 32]$ &  \\
        & $f(\vec{x}^*)$ = 0, with $x^* = (0, \ldots, 0)$ &  \\
        \hline
        
          \multirow{3}{*}{Eggholder}  & $ f(\vec{x}) = \sum_{i=1}^{n-1}[-x_i sin(\sqrt{|x_i - x_{i+1} -47|}) - (x_{i+1}+47) sin(\sqrt{|0.5 x_i + x_{i+1} +47|})]$ & \multirow{3}{*}{ \includegraphics[ height=4.1em]{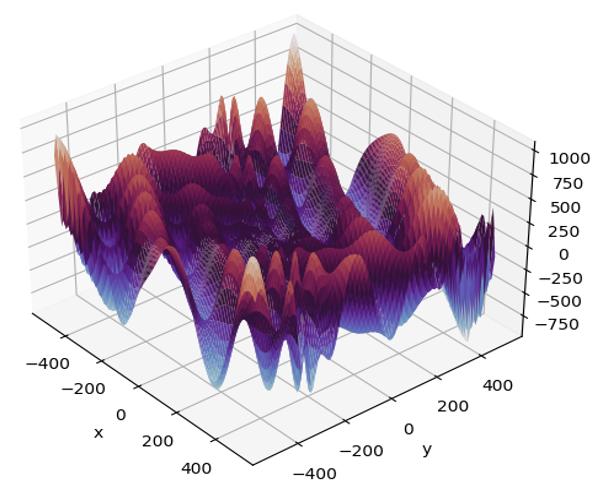}}\\
        & $x_i \in [-512, 512]$ &  \\
        & $f(\vec{x}^*) = -959.64$, with $x^* = (512, 404)$ for $n=2$ &  \\
        \hline
        
         \multirow{3}{*}{Griewank}  & $ f(\vec{x}) = 1 + \sum_{i=1}^{n} \frac{x_i^{2}}{4000} - \prod_{i=1}^{n}cos(\frac{x_i}{\sqrt{i}})$ & \multirow{3}{*}{ \includegraphics[ height=4.1em]{figures/functions/griewank.png}}\\
        & $x_i \in [-600, 600]$ &  \\
        & $f(\vec{x}^*)$ = 0, with $x^* = (0, \ldots , 0)$ &  \\
        \hline
        
         \multirow{3}{*}{Happy Cat}  & $ f(\vec{x}) =\left[\left(||\textbf{x}||^2 - n\right)^2\right]^\alpha + \frac{1}{n}\left(\frac{1}{2}||\textbf{x}||^2+\sum_{i=1}^{n}x_i\right)+\frac{1}{2}$ & \multirow{3}{*}{ \includegraphics[ height=4.1em]{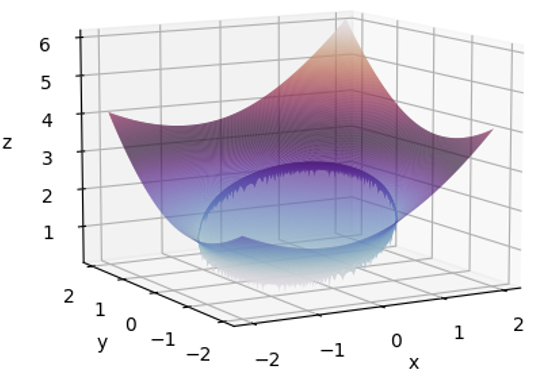}}\\
        & $x_i \in [-2, 2]$ &  \\
        & $f(\vec{x}^*)$ = 0, with $x^* = (-1, \ldots , -1)$ &  \\
        \hline
        
        \multirow{4}{*}{Rana}  & $ f(\vec{x}) = \sum_{i=1}^n [x_i sin (t_2) cos(t_1) + (x_1 + 1)sin(t_1) cos(t_2) ]$ & \multirow{4}{*}[-0.5ex]{ \includegraphics[ height=4.1em]{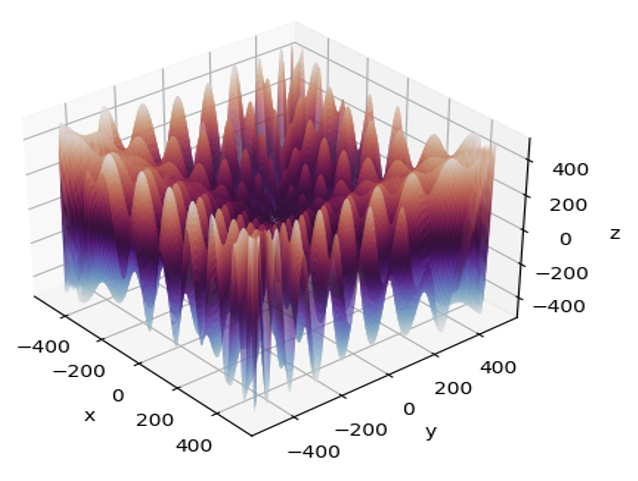}}\\
        & with $t_1 = \sqrt{|x_1+x_i+1|}$ and $t_2 = \sqrt{|x_1 - x_i +1|}$ & \\
        & $x_i \in [-500, 500]$ &  \\
        & $f(\vec{x}^*) = -959.64$, with $x^* = (512, 404)$ for $n=2$ &  \\
        \hline
        
          \multirow{3}{*}{Rastrigin}  & $ f(\vec{x}) = 10n + \sum_{i=1}^{n}(x_i^2 - 10cos(2\pi x_i))$ & \multirow{3}{*}{ \includegraphics[ height=4.1em]{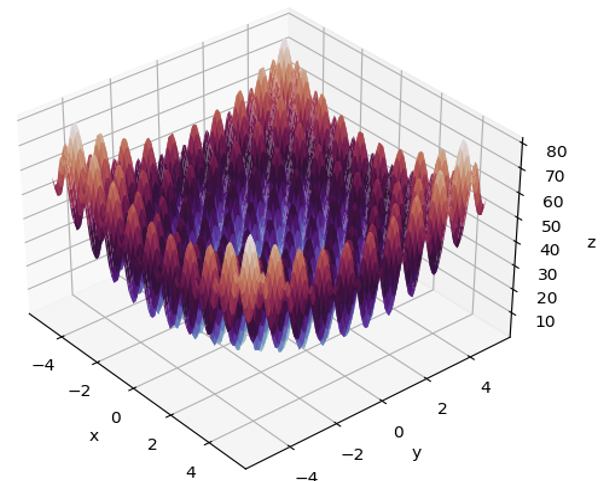}}\\
        & $x_i \in [-5.12, 5.12]$ &  \\
        & $f(\vec{x}^*)$ = 0, with $x^* = (0, \ldots, 0)$ &  \\
        \hline
        
         \multirow{3}{*}{Rosenbrock}  & $ f(\vec{x}) =\sum_{i=1}^{n}[100 (x_{i+1} - x_i^2)^ 2 + (1 - x_i)^2]$ & \multirow{3}{*}{ \includegraphics[ height=4.1em]{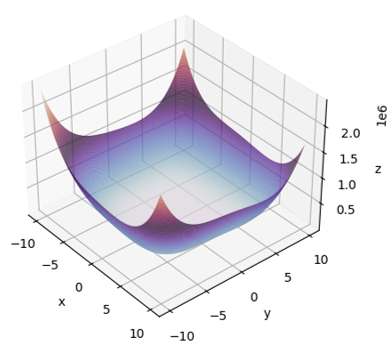}}\\
        & $x_i \in [-5, 10]$ &  \\
        & $f(\vec{x}^*)$ = 0, with $x^* = (1,\ldots, 1)$ &  \\
        \hline
        
          \multirow{3}{*}{Salomon}  & $ f(\vec{x}) = 1-cos(2\pi\sqrt{\sum_{i=1}^{D}x_i^2})+0.1\sqrt{\sum_{i=1}^{D}x_i^2}$ & \multirow{3}{*}{ \includegraphics[ height=4.1em]{figures/functions/salomon_glob.png}}\\
        & $x_i \in [-100, 100]$ &  \\
        & $f(\vec{x}^*)$ = 0, with $x^* = (0, \ldots , 0)$ &  \\
        \hline
        
         \multirow{3}{*}{Sargan}  & $ f(\vec{x}) = \sum_{i=1}^n n (x_i^2 + 0.4 \sum_{j \neq i}^n x_i x_j)$ & \multirow{3}{*}{ \includegraphics[ height=4.1em]{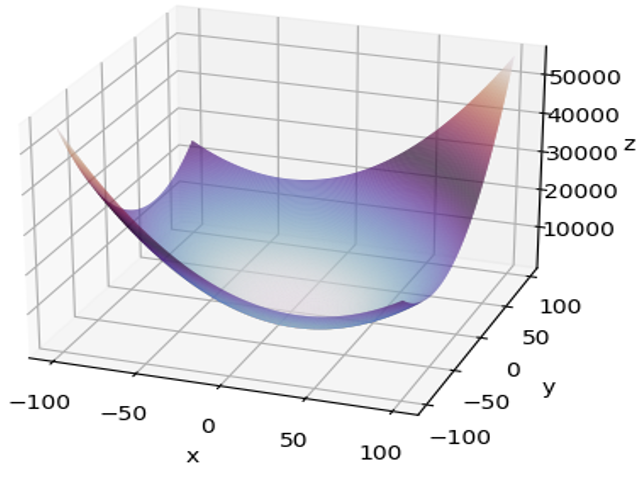}}\\
        & $x_i \in [-100, 100]$ &  \\
        & $f(\vec{x}^*)$ = 0, with $x^* = (0, \ldots , 0)$ &  \\
        \hline
        
          \multirow{3}{*}{Schaffer F6}  & $ f(\vec{x}) = \sum_{i=1}^n 0.5+\frac{sin^2(\sqrt{x_i^2 + x_{i+1}^2})-0.5}{[1+0.001 \cdot (x_i^2 + x_{i+1}^2)]^2}$ & \multirow{3}{*}{ \includegraphics[ height=4.1em]{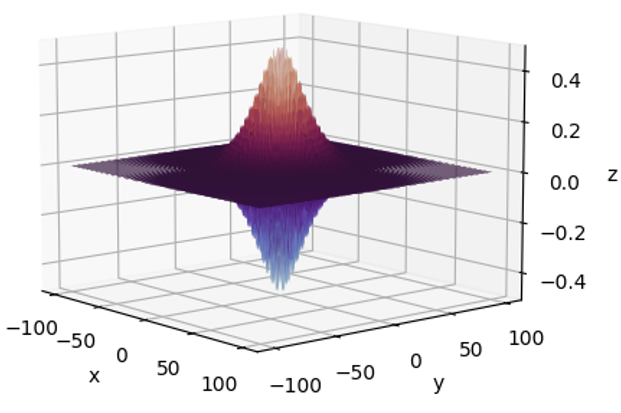}}\\
        & $x_i \in [-100, 100]$ &  \\
        & $f(\vec{x}^*)$ = 0, with $x^* = (0, \ldots , 0)$ &  \\
        \hline
        
        \multirow{3}{*}{Schwefel 2.26}  & $ f(\vec{x}) = 418.9829 n - \sum \limits_{i=1}^n x_i \sin{\sqrt{|x_i|}}$ & \multirow{3}{*}[0.5ex]{ \includegraphics[width=0.14\textwidth]{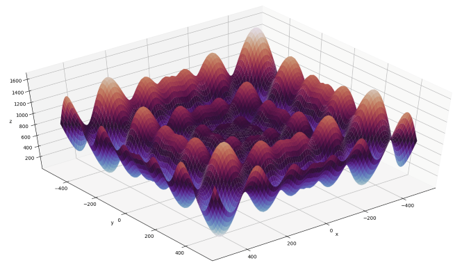}}\\
        & $x_i \in [-500, 500]$ &  \\
        & $f(\vec{x}^*)$ = 0, with $x^* = (420,968746, \ldots, 420,968746)$ &  \\
        \hline
        
        \multirow{3}{*}{Penalized Schwefel 2.26} & $ f(\vec{x}) = 418.9829 n - \sum \limits_{i=1}^n x_i \sin{\sqrt{|x_i|}} + \biggl| \: \sum \limits_{i=1}^{\frac{n}{2}} x_{2i} - \sum \limits_{i=1}^{\frac{n}{2}} x_{2i-1} \: \biggl|$ & \multirow{3}{*}[0.5ex]{ \includegraphics[width=0.14\textwidth]{figures/functions/schwefel_penalty.png}} \\
          & $x_i \in [-500, 500]$ &  \\
        & $f(\vec{x}^*)$ = 0, with $x^* = (420,968746, \ldots, 420,968746)$ &  \\
          \hline
          
         \multirow{3}{*}{Qing}  & $ f(\vec{x}) = \sum_{i=1}^{n}(x^2-i)^2$ & \multirow{3}{*}{ \includegraphics[ height=4.1em]{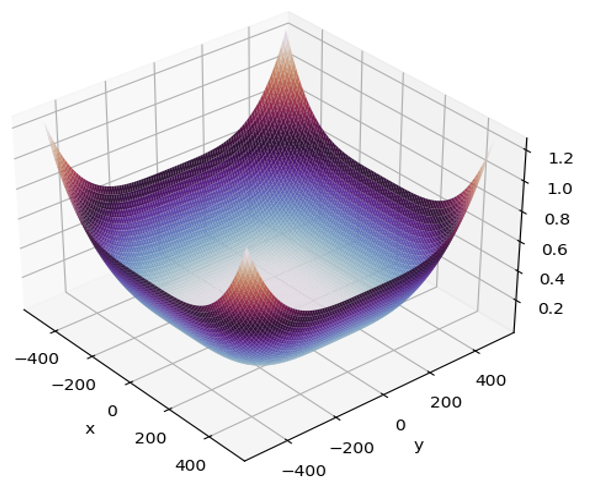}}\\
        & $x_i \in [-500, 500]$ &  \\
        & $f(\vec{x}^*)$ = 0, with $x^* = ( \pm \sqrt{i}, \ldots , \pm \sqrt{i})$ &  \\
        \hline
   \end{tabularx}
    \caption{Benchmark functions}
    \label{tab:benchmark_functions}
\end{table*}

\begin{table*}[ht]
    \centering
\begin{tabular}{ |c|c|c|c|c|c|c|c|c|c| }
\hline
& dimension & $DM$  & $FEM_{macro}$  & $FEM_{micro}$  & $SEM_{macro}$  & $SEM_{micro}$  & $PIC_{macro}$  & $PIC_{micro}$  & $separable$ \\
\hline
\multirow{2}{*}{$schwefel_{2.2.6} \, Pen$} & 50  & -0.006  & 0.869  & 0.659  & 0.566  & 0.729  & 0.609  & 0.316  & 0  \\
 & 100  & -0.005  & 0.87  & 0.667  & 0.57  & 0.733  & 0.61  & 0.324  & 0 \\
\hline
\multirow{2}{*}{$schwefel_{2.2.6}$} & 50  & 0.011  & 0.873  & 0.667  & 0.56  & 0.731  & 0.615  & 0.316  & 1  \\
 & 100  & -0.001  & 0.874  & 0.668  & 0.565  & 0.734  & 0.622  & 0.326  & 1 \\
\hline
\multirow{2}{*}{$rastrigin$} & 50  & -0.278  & 0.881  & 0.641  & 0.522  & 0.722  & 0.658  & 0.298  & 1  \\
 & 100  & -0.191  & 0.88  & 0.62  & 0.526  & 0.728  & 0.655  & 0.278  & 1 \\
\hline
\multirow{2}{*}{$griewank$} & 50  & -0.395  & 0.695  & 0.336  & 0.703  & 0.798  & 0.354  & 0.101  & 0  \\
 & 100  & -0.273  & 0.694  & 0.281  & 0.707  & 0.804  & 0.349  & 0.07  & 0 \\
\hline
\multirow{2}{*}{$ackley$} & 50  & -0.322  & 0.872  & 0.885  & 0.531  & 0.515  & 0.64  & 0.682  & 0  \\
 & 100  & -0.216  & 0.868  & 0.884  & 0.536  & 0.526  & 0.642  & 0.665  & 0 \\
\hline
\multirow{2}{*}{$eggholder$} & 50  & 0.053  & 0.882  & 0.733  & 0.527  & 0.677  & 0.659  & 0.4  & 0  \\
 & 100  & 0.025  & 0.884  & 0.739  & 0.531  & 0.677  & 0.664  & 0.406  & 0 \\
\hline
\multirow{2}{*}{$qing$} & 50  & -0.364  & 0.747  & 0.398  & 0.667  & 0.788  & 0.405  & 0.138  & 0  \\
 & 100  & -0.254  & 0.743  & 0.351  & 0.674  & 0.796  & 0.407  & 0.118  & 0 \\
\hline
\multirow{2}{*}{$salomon$} & 50  & -0.372  & 0.83  & 0.896  & 0.605  & 0.519  & 0.526  & 0.643  & 0  \\
 & 100  & -0.262  & 0.816  & 0.881  & 0.614  & 0.54  & 0.523  & 0.611  & 0 \\
\hline
\multirow{2}{*}{$rana$} & 50  & 0.024  & 0.887  & 0.771  & 0.505  & 0.653  & 0.698  & 0.453  & 0  \\
 & 100  & 0.014  & 0.889  & 0.776  & 0.508  & 0.652  & 0.7  & 0.463  & 0 \\
\hline
\multirow{2}{*}{$sargan$} & 50  & -0.28  & 0.69  & 0.603  & 0.728  & 0.783  & 0.357  & 0.265  & 0  \\
 & 100  & -0.173  & 0.697  & 0.618  & 0.735  & 0.79  & 0.373  & 0.28  & 0 \\
\hline
\multirow{2}{*}{$schaffer_{f6}$} & 50  & -0.066  & 0.875  & 0.755  & 0.509  & 0.568  & 0.693  & 0.737  & 0  \\
 & 100  & -0.038  & 0.882  & 0.758  & 0.51  & 0.572  & 0.698  & 0.737  & 0 \\
\hline
\multirow{2}{*}{$rosenbrock$} & 50  & -0.291  & 0.727  & 0.4  & 0.678  & 0.784  & 0.382  & 0.135  & 0  \\
 & 100  & -0.192  & 0.722  & 0.343  & 0.684  & 0.795  & 0.381  & 0.109  & 0 \\
\hline
\multirow{2}{*}{$happy cat$} & 50  & -0.374  & 0.697  & 0.241  & 0.696  & 0.773  & 0.355  & 0.115  & 0  \\
 & 100  & -0.263  & 0.688  & 0.197  & 0.705  & 0.777  & 0.353  & 0.085  & 0 \\
\hline
\end{tabular}
    \caption{Fitness landscape metrics for reflected benchmarks}
    \label{tab:func_metrics}
\end{table*}

\end{document}